\pdfoutput=1
\documentclass[10pt,twocolumn,superscriptaddress,aps,prd,nofootinbib]{revtex4-2}
\usepackage{graphicx}
\usepackage{subfigure}
\usepackage{dcolumn}\usepackage{bm}
\usepackage[usenames]{color}
\usepackage{amsmath}
\usepackage{amssymb}
\usepackage{url} 
\usepackage{hyperref} 
\usepackage{amsmath}
\usepackage{accents}
\usepackage{lineno}
\usepackage{multirow}

\newcommand*{\Apr}{{A^\prime}}

\newcommand{\beq}{\begin{equation}}
\newcommand{\eeq}{\end{equation}}
\begin{document}
\title{Dark matter search with the BDX-MINI experiment}  

\newcommand*{\JLAB}{Thomas Jefferson National Accelerator Facility, 23606, Newport News, VA, USA}

\newcommand*{\INFNGE}{Istituto Nazionale di Fisica Nucleare, Sezione di Genova, 16146 Genova, Italy}

\newcommand*{\INFNCT}{Istituto Nazionale di Fisica Nucleare, Sezione di Catania, 95125 Catania, Italy}

\newcommand*{\UNIGE}{Universit\'a degli Studi di Genova, 16126 Genova, Italy}
\newcommand*{\LAMAR}{Lamar University, 4400 MLK Boulevard, P.O. Box 10046, Beaumont, Texas 77710, USA}
\newcommand*{\OXY}{Occidental College, Los Angeles, CA 90041, USA }
\newcommand*{\CANISIUS}{Canisius College, Buffalo, NY, 14208, USA}
\newcommand*{\INFNRM}{Istituto Nazionale di Fisica Nucleare, Sezione di Roma2, 00133 Roma, Italy}
\newcommand*{\UNITV}{Universit\'a di Roma Tor Vergata, 00133 Roma, Italy }

\author {M.~Battaglieri} 
\affiliation{\JLAB}
\affiliation{\INFNGE}
\author {M.~Bond\'i}
\affiliation{\INFNRM}
\affiliation{\UNITV}
%\author {P.~Bisio} 
%\affiliation{\UNIGE}
\author {A.~Celentano} 
\affiliation{\INFNGE}
\author {P.L. Cole} 
\affiliation{\LAMAR}
\author {M.~De Napoli} 
\affiliation{\INFNCT}
\author {R.~De Vita} 
\affiliation{\INFNGE}
\author {L.~Marsicano}
\email[Corresponding author:]{luca.marsicano@ge.infn.it}
\affiliation{\INFNGE}
\author {N.~Randazzo} 
\affiliation{\INFNCT}
\author {E.S. Smith} 
\affiliation{\JLAB}
\author {D.~Snowden-Ifft} 
\affiliation{\OXY}
\author {M.~Spreafico} 
\affiliation{\UNIGE}
\affiliation{\INFNGE}
\author {M.H.~Wood} 
\affiliation{\CANISIUS}

\date{\today}
\begin{abstract}
BDX-MINI is a beam dump experiment optimized to search for Light Dark Matter 
produced in the interaction of the intense CEBAF 2.176 GeV electron beam with the  Hall A beam dump at Jefferson Lab. 
The BDX-MINI detector consists of a PbWO$_4$ electromagnetic calorimeter surrounded by a hermetic veto system for background rejection. The experiment accumulated $2.56 \times 10^{21}$ EOT in six months of running. Simulations of fermionic and scalar Dark Matter interactions with electrons of the active volume of the BDX-MINI detector were used to estimate the expected signal.
Data collected during the beam-off time allowed us to characterize the background dominated by cosmic rays. A blind data analysis based on a maximum-likelihood approach was used to optimize the experiment sensitivity. An upper limit on the production of light dark matter was set using the combined event samples collected during beam-on and beam-off configurations. In some kinematics, this pilot experiment is sensitive to the parameter space covered by some of the most sensitive experiments to date, which demonstrates the discovery potential of the next generation beam dump experiment planned at intense electron beam facilities.
\end{abstract}

\pacs{123456789} 
\maketitle
\section{\label{sec:intro} Introduction and Theoretical framework}

Many astrophysical observations as well as anomalies in processes involving electromagnetic currents (e.g. the muon anomalous magnetic moment) could be reconciled assuming the existence of a new kind of matter, not directly interacting with light, called {\it Dark Matter} (DM)~\cite{Cebrian:2022brv,Arbey:2021gdg}. While
gravitational effects of DM are quite well established, despite the tremendous efforts being devoted to reveal the nature of DM in terms of new elementary particles, no clear results have been obtained to date. Experimental efforts have mainly focused on direct detection of galactic DM  within the so called ``WIMPs paradigm'' that assumes the existence of slow-moving cosmological weakly interacting particles with mass larger than 1 GeV~\cite{STEIGMAN1985375,Roszkowski:2017nbc,Schumann:2019eaa}. 
Due to the lack of evidence in ``traditional'' DM searches, there has been increased experimental activity directed toward the search for  light DM (LDM) in the MeV-GeV mass range~\cite{Battaglieri:2017aum, Krnjaic:2022ozp,Fabbrichesi:2020wbt,Filippi:2020kii}. This largely unexplored mass region is theoretically well justified with the assumption that DM has a thermal origin~\cite{Liddle:1998ew,Coy:2021ann}.
If dark and visible matter  had sufficiently large interactions to 
achieve thermal equilibrium during the early phases of the nascent universe, due to the Hubble expansion that diluted the DM number density, the annihilation rate correspondingly lessened becoming fixed for all subsequent times thereafter. Therefore, in this hypothesis there has been a sufficiently high DM annihilation rate to have
depleted any excess abundance in order to agree with our present day observations.
For thermal dark matter below the GeV scale, this requirement can only be satisfied if the dark sector contains comparably light new force carriers to mediate the necessary annihilation process. Such mediators must couple to visible matter and neutral under the Standard Model (SM) gauge group. This greatly limits the number of options for possible mediators. A representative model involves a heavy vector boson called $\Apr$ or ``heavy photon''~\cite{Holdom:1985ag, Batell:2009di,Fabbrichesi:2020wbt} and is described by the Lagrangian (after fields diagonalization, and omitting the LDM mass term):
\begin{equation}\label{eq:lagrangian}
%\begin{split}
 \mathcal{L} \supset -\frac{1}{4} F'_{\mu\nu}F'^{\mu\nu}
 +\frac{1}{2}m^2_\Apr \Apr_\mu \Apr^\mu
 -\frac{\varepsilon}{2}F_{\mu\nu}F'^{\mu\nu} -g_D \Apr_\mu J^\mu_D 
%\end{split}
\end{equation}
where $m_\Apr$ is the dark photon mass, $F'_{\mu\nu}\equiv \partial_\mu A'_\nu - \partial_\nu A'_\mu$ is the dark photon field strength, $F_{\mu\nu}$ is the SM electromagnetic field strength, $g_D\equiv\sqrt{4 \pi \alpha_D}$ is the dark gauge coupling, $J^\mu_D$ is the current of DM fields and $\varepsilon$ parametrizes the degree of kinetic mixing between dark and visible photons. The phenomenology of the DM interaction depends on the DM mediator mass hierarchy and on the details of the dark current $J^\mu_D$. If there is only one dark sector state, the dark current generically contains elastic interactions with the
dark photon. However, if there are two (or more) dark sector states the dark photon can couple to
the dark sector states off-diagonally~\cite{Tucker-Smith:2001myb}. In this work, we will consider the two cases of scalar and fermionic LDM; although the latter has been already strongly constrained by CMB arguments~\cite{Madhavacheril:2013cna}, it is representative of other model variations, such as the Majorana or the pseudo-Dirac (small mass splitting) cases~\cite{LDMX:2018cma}. Depending on the relative masses of the $\Apr$ and the DM particles $\chi$ the $\Apr$ can  decay to SM particles (``visible'' decay) and/or to light DM states (``invisible'' decay). In the rest of the paper we will consider the mass hierarchy case  $m_{\Apr} > m_\chi$ with the arbitrary choice $m_{\Apr}/m_\chi$=3, where $m_\chi$ is the LDM mass. We will also fix $\alpha_D=0.1$, following the recent convention adopted in the CERN Physics Beyond Collider report~\cite{Beacham:2019nyx}.

In the paradigm of DM with a thermal origin, it would have acquired its current abundance
through direct or indirect annihilation into SM. If the mediator is heavier than the DM, the thermal
relic abundance is achieved via direct annihilation $\chi \chi \to f f$ where $f$ are SM fermions/scalars with a corresponding annihilation rate scaling as:

\begin{equation}
    \sigma v_{\chi \chi \to f f} \propto y \equiv \varepsilon^2 \alpha_D (m_\chi / m_{A'})^4.
\end{equation}
This scenario offers a predictive target for discovery or falsifiability, since there is a minimum
SM-mediator coupling compatible with a thermal history that experiments can probe.

Present accelerator technology provides high intensity particle beams of moderate energy that are well suited for the discovery of LDM~\cite{Battaglieri:2017aum, Krnjaic:2022ozp}.  In particular, electron beam dump experiments have been shown to have high sensitivity to light dark matter~\cite{Batell:2014mga, Andreev:2021fzd, Bondi:2017gul}. In these experiments, light dark matter particles ($\chi {\bar\chi}$ pairs) are conjectured to be produced  when the electron
beam interacts with nucleons and electrons  in the beam dump via $\Apr$ radiative process ($\Apr$-\textit{strahlung}) or annihilation (resonant and non-resonant) of positrons produced by the electromagnetic shower generated therein~\cite{Izaguirre:2013uxa, Marsicano:2018glj, Marsicano:2018krp}. A detector located downstream of the beam dump, shielded from SM background particles (other than neutrinos), could be sensitive to the interaction of DM $\chi$s. However, the shallow installation of accelerator beam lines exposes detectors to cosmic backgrounds~\cite{BDX:2019afh}. 
An electromagnetic shower is expected to be produced by the $\chi$-electron interaction, depositing large energies (E$_{Dep}>10$ MeV) that are easily detectable by a standard electromagnetic calorimeter. Surrounding the detector with passive and active vetoes provide further reduction of beam-related  and cosmic muon and neutron backgrounds.

An experiment based on the concept described above, called  Beam Dump eXperiment or BDX~\cite{Bondi:2017gul}, has been proposed and approved to run at Jefferson Lab making use of the 11 GeV, 60 $\mu$A CW electron beam delivered  by the CEBAF accelerator to the experimental Hall A. Five meters of iron will shield a cubic-meter-size electromagnetic calorimeter consisting of  CsI(Tl) crystals formerly used in the BaBar experiment at SLAC~\cite{BaBar:2001yhh}. Plastic scintillators and lead layers surrounding the calorimeter will veto most of the cosmic and beam-related backgrounds.

In this paper we describe the  BDX-MINI experiment~\cite{Battaglieri:2020lds}, a pilot version of BDX that took data at lower beam energy (2 GeV in place of 11 GeV), and therefore avoided the need for the significant shielding planned for the full experiment.
BDX-MINI was installed in an existing 10-inch pipe installed 25~m downstream of the Hall A beam dump to intercept possible dark matter particles propagating through the dirt. To partially compensate for the limited active volume that was constrained by the pipe size, the calorimeter was assembled using $\rm PbWO_4$ crystals, which are a factor of two denser than CsI(Tl). Following the BDX detector design, two layers of plastic scintillator and a passive layer of W surrounded the calorimeter. While the BDX-MINI interaction volume was only a few percent of proposed BDX and the beam exposure was only six months, the outcome of this pilot experiment is worthy in its own right. 

The paper is organized as follow: Sec.~\ref{sec:setup} describes the experimental set up: location, detector, and  DAQ. Sec.~\ref{sec:simulation} describes the simulation framework used to evaluate the expected signal. Section~\ref{sec:analysis} reports details about the data analysis and the statistical procedure.
Finally, Sec.~\ref{sec:results} reports the experiment sensitivity optimization and the resulting upper limit. A Summary and Outlook will conclude the paper.

%\section{\label{sec:setup} Experimental setup}
\section{\label{sec:setup} Experimental setup}
\begin{figure}[t]
\centering
\includegraphics[width=2.9in]{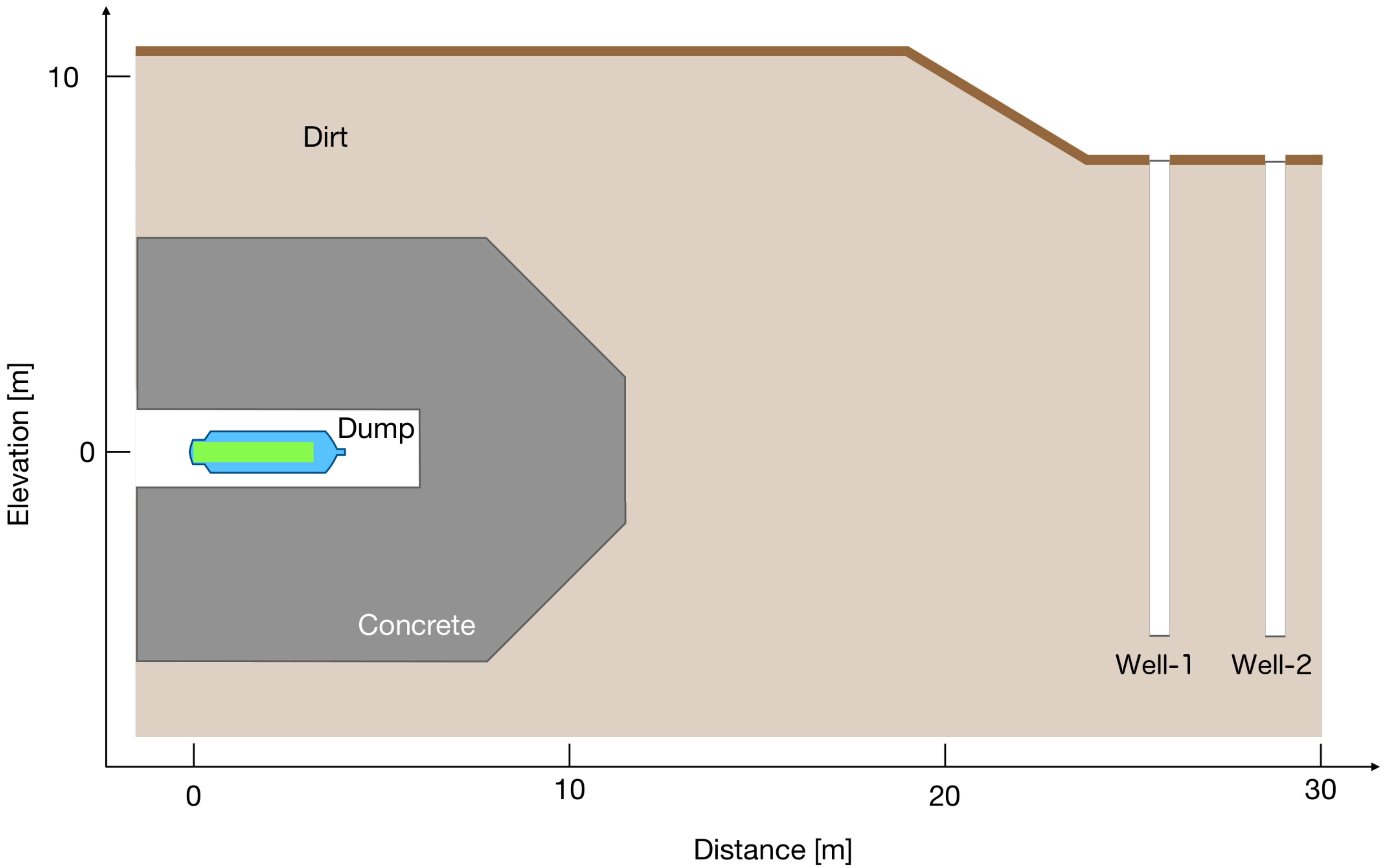}
\caption{Schematic representation of the location of the wells relative to the Hall A beam dump. From left to right, the Hall A aluminum-water beam dump (blue-green), the concrete beam vault walls (gray), the dirt (brown), and the two vertical pipes. The detector was located in the well closest to the accelerator, Well-1. (Color online)}
 \label{fig:pipes_location}
\end{figure}

%Initial draft 1/28/2022  ES
The detector was installed  26 m downstream of the Hall A beam dump at JLab, inside a well (Well-1) at beamline height.
The detector was shielded from the  background produced  in the beam dump by 5.4 m of concrete and 14.2 m of dirt, shown schematically in Fig.\,\ref{fig:pipes_location}. The experiment accumulated data during a period of six months between 2019 and 2020. During most of the time, Hall A received one-pass beam from the accelerator (2.176~GeV) and all muons generated in the beam dump ranged out before reaching the detector. In this section we present an 
overview of the experimental configuration and detector with sufficient information to understand the analysis and results in the rest of the paper. 
Additional details of the experimental layout and setup can be found in Ref.~\cite{Battaglieri:2020lds}. 

The detector was located inside a 20-cm diameter stainless steel watertight cylindrical vessel, which could be lowered into either one of the two wells downstream of the Hall A beam dump. Cable connections from the detector were routed to an electronic rack at ground level near the entrance to the well that contained the readout electronics and the DAQ system. The experimental setup was housed inside a sturdy field tent that covered both wells, the electronic equipment, and power breakers. The environment was conditioned using a portable air conditioning unit to maintain a suitable temperature and humidity.

\begin{figure}[t]
\centering
\includegraphics[width=.45\textwidth]{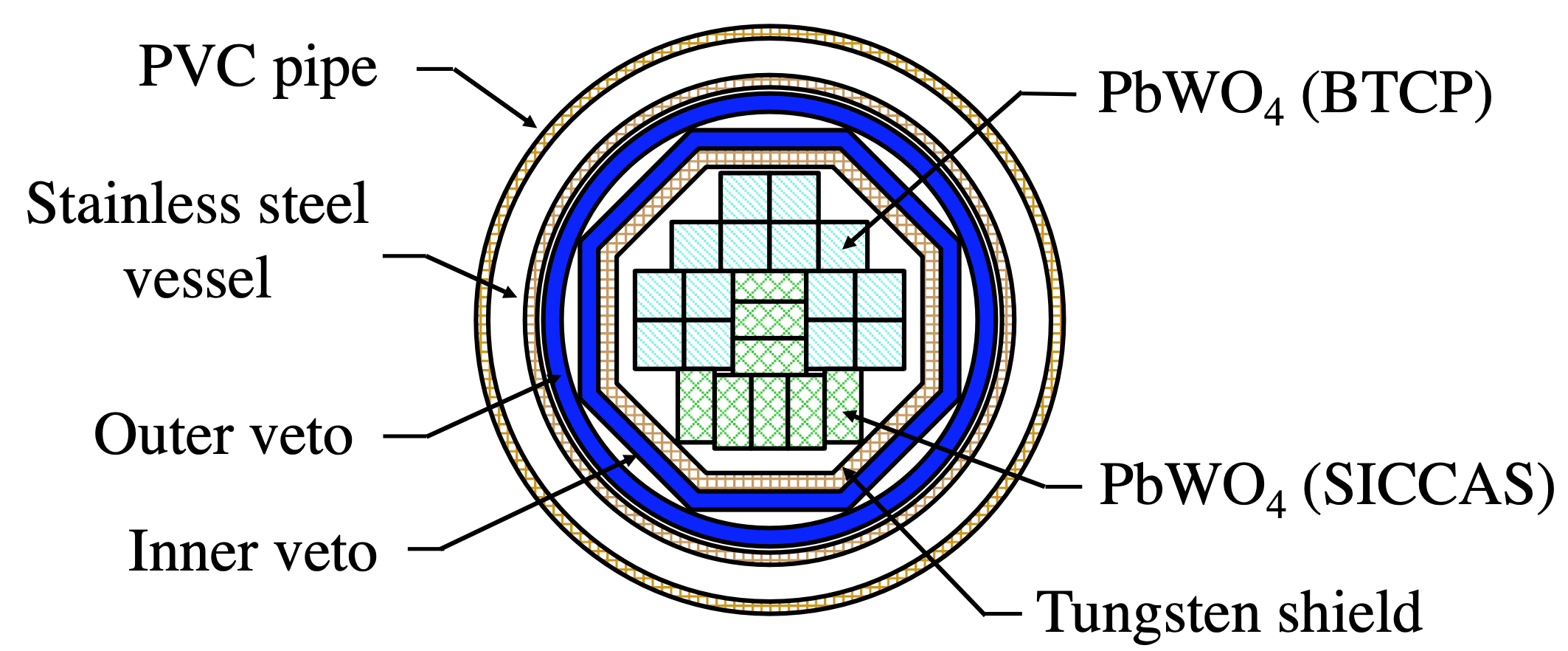}
\caption{Sketch of the ``BDX-MINI'' detector. The detector is located inside a well, illustrated in Fig.\,\ref{fig:pipes_location}, whose wall is shown as the outer PVC cylinder in this sketch. The detector package fits inside the stainless steel vessel.
 \label{fig:detector_sketch}}
\end{figure}

The detector package, which we call ``BDX-MINI'', consists of an electromagnetic calorimeter (ECal) composed of PbWO$_{4}$ crystals ($\sim$\,4\,x\,10$^{-3}$\,m$^{3}$ total volume) and hermetic layers of passive and active vetoes. 
The innermost veto layer consists of tungsten shielding. This passive layer is followed by two active layers: the inner veto, IV, and the outer veto, OV. Each layer consists of a cylindrical or octagonal tube and two end caps. A cross sectional sketch of the detector is shown in Fig.\,\ref{fig:detector_sketch}. 

The ECal is composed of two identical modules with 22 PbWO$_{4}$ crystals each (eight 30\,x\,15\,x\,200\,mm$^{3}$ crystals produced by SICCAS\footnote{Purchased for the Forward Tagger Detector for CLAS12 from Shanghai Institute of Ceramics, Chinese Academy of Science.} and fourteen 20\,x\,20\,x\,200\,mm$^{3}$ crystals produced by the BTCP\footnote{Purchased for the PANDA-ECal from Bogoroditsk Technical Chemical Plant.}). The two modules are mounted vertically, resulting in an approximately cylindrical shape 40\,cm long and 11.5~cm in diameter (equivalent to approximately 13 radiation lengths). The ECal uses 6\,x\,6\,mm$^{2}$ Hamamatsu MPPCs (S13360-6025PE) to read out the PbWO$_{4}$ scintillation light from each crystal.

The veto is composed of three layers, which completely enclose the ECal. The innermost passive layer is a  0.8\,cm-thick tungsten shield that is shaped as an octagonal prism with a height of 45\,cm and a base side of 5\,cm. This layer is sealed on top and bottom by two octagonal tungsten plates. The purpose of this passive layer is to protect the active vetoes from electromagnetic showers in the ECAL produced by $\chi - e^-$ interactions that may accidentally self-veto the interaction. The two active layers are composed of EJ200 plastic scintillators, 0.8\,cm thick. The IV is an optically connected octagonal prism with a side base of 6.2\,cm and a height of 49.4\,cm; two octagonal scintillators cover the top and bottom. The OV is cylindrical in shape with a radius of 9.7\,cm and a height of 53.0\,cm; the top and bottom are covered by two round caps. Both the IV and the OV use wavelength-shifting fibers to collect the light and deliver it to 3x3\,mm$^{2}$ Hamamatsu S13360-3075CS SiPMs with multiple redundancy. 

The preamplifiers, readout electronics and data acquisition electronics are all mounted in racks outside the well at ground level.
We found that the noise and the attenuation associated to 8-m long cables between the SiPMs and the front-end electronics were sufficiently low to distinguish the individual photo-electrons signals for each and every one of the SiPMs. Each detector signal is amplified by a custom circuit, which provides two equal outputs of the signal. The first copy of the signal is sent to a leading-edge discriminator (CAEN v895), whose digital output is sent to a programmable logic board (CAEN FPGA v1495) implementing a custom trigger logic. A threshold of $\sim10$ MeV is implemented for each crystal, while for veto signals a threshold of few photo-electrons is used.
The second copy is fed to a Flash Amplitude-to-Digital converter. Signals from the inner and outer veto photodetectors were processed with a 2V, 14 bit, 500 MHz module (CAEN FADC v1730), while those from the crystals were processed with a board featuring a lower 250 MHz sampling rate (CAEN FADC v1725). To guarantee synchronization, the clock was generated from the first board and distributed to the others through a daisy chain setup. A 640~ns readout window was used for all channels.

The main experimental trigger consisted of the logic ``OR'' of all signals from the crystals in the ECal, resulting in a constant rate of 3.2\,Hz. The rate was insensitive to the beam operation. Other triggers were implemented for monitoring, calibration and debugging. Each individual trigger could be prescaled, and the global trigger condition consisted in the union of all individual trigger bits after pre\-scale. For each trigger, all the FADC raw waveforms were written to the disk without further processing. In order to monitor the rates in the detector, as well as the trigger rate and the livetime, individual scalers were implemented in the FPGA firmware and regularly read through the slow controls system. We used the standard JLab ``CEBAF Online Data Acquisition'' (CODA) software to handle the readout system~\cite{CODA}.

A custom EPICS-based system was developed~\cite{Dalesio:1994qp}, which was integrated into the main JLab slow-controls in order to access accelerator quantities such as the beam current and energy.  The temperature and the humidity at the detector was monitored by two probes installed inside the watertight cylindrical vessel. Likewise, the ambient temperature and humidity inside the tent with the electronics were recorded. The FPGA scalers were readout and trigger configurations could be set using the slow-control system. All slow-control variables were periodically recorded to the data stream during data-taking runs.

Various online monitoring tools were developed to monitor the detector performance during data taking in order to quickly identify and thereby correct any problems as quickly as possible. The count rates for each detector channel, for each trigger bit, and for the total trigger rate were available as data came in and their time evolution was monitored using the EPICS StripTool program. The online reconstruction system also monitored both single-event FADC waveforms and accumulated observables, including the spectrum of the energy deposited in each crystal, and the total energy deposited in the two BDX-MINI modules. If anomalies were found by any of these tools, the DAQ was suspended and the problem investigated till resolved.

\section{\label{sec:simulation} Expected Signal and Simulation}
% \subsection{\label{sec:signal} Signal yield}
The sensitivity of the BDX-MINI experiment to the production of LDM is determined by precisely simulating the production of LDM in the dump and the response of the detector to interactions  of crossing LDM particles. The distributions are computed for specific models and masses of LDM. The expectation for the signal is then compared to the measured distributions in our data to determine, or set limits on, the production strength of LDM.
In the following, we first summarize the relevant formulas for LDM production in the beam dump and detection in BDX-MINI, and then we present the Monte Carlo strategy that we adopted to compute the expected signal yield.

\subsection{LDM production and detection}

In BDX-MINI, LDM particles are produced by the interaction of the secondary particles in the electromagnetic shower induced by the CEBAF electron beam with the nuclei and electrons contained in the Hall A beam dump.
The main production mechanisms are the so-called $\Apr$-strahlung ($e^\pm N \rightarrow e^\pm N \Apr$) and the resonant $e^+e^-$ annihilation ($e^+e^-\rightarrow\Apr$), followed by the invisible $\Apr\rightarrow \chi\overline{\chi}$ decay to LDM particles~\cite{Marsicano:2018glj,Izaguirre:2013uxa}. The $\chi$ and $\overline{\chi}$ particles are sufficiently long lived and penetrating to reach, and possibly interact with, the BDX-MINI detector.

\begin{figure}[t]
    \centering
    \includegraphics[width=.48\textwidth]{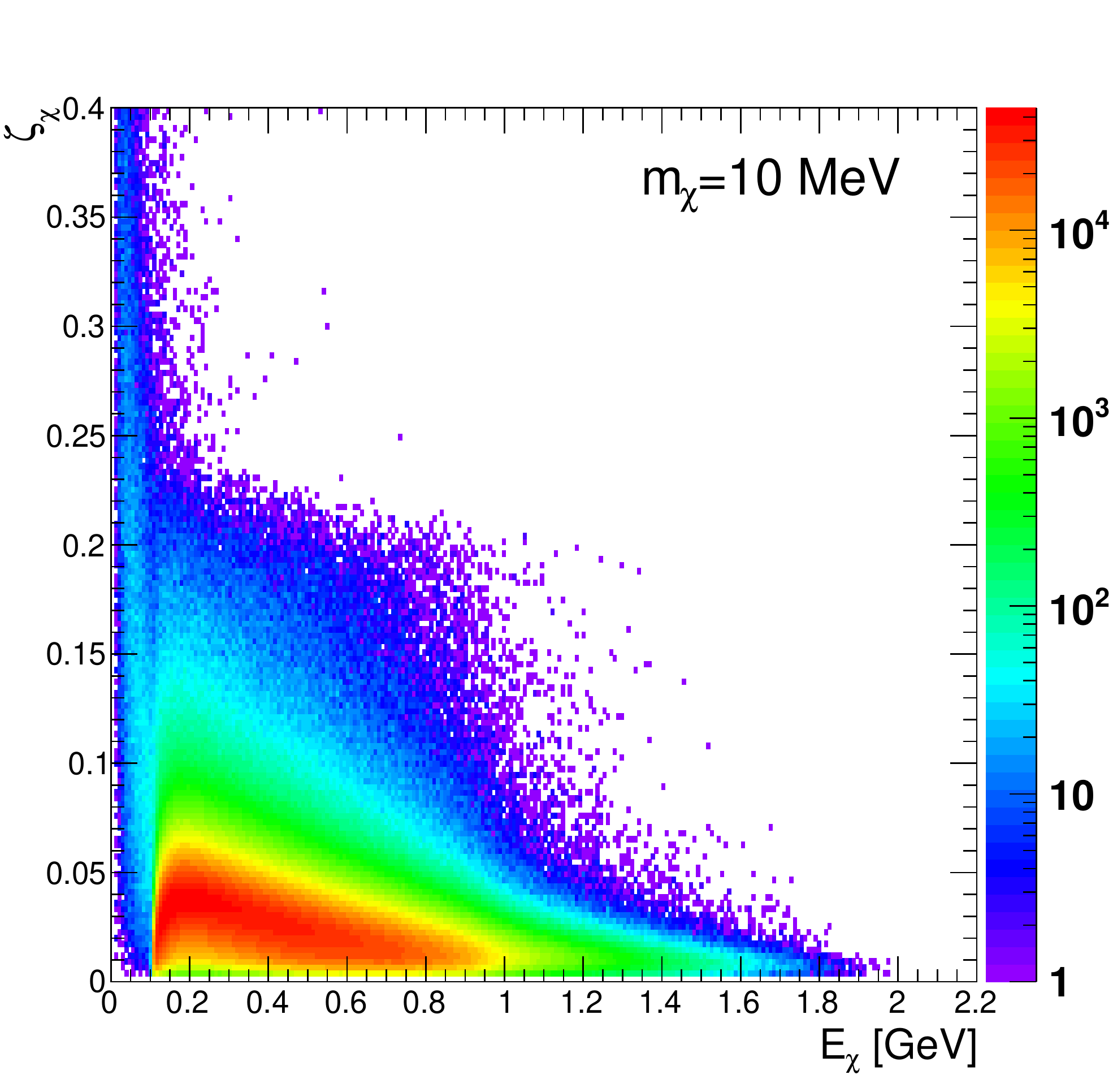}
    \caption{LDM particle flux (in arbitrary units) downstream of the Hall A beam dump for fermionic LDM of $m_\chi=10$ MeV, as a function of the particle energy $E_\chi$ and polar angle $\zeta_\chi$.}
    \label{fig:ChiSpectrum}
\end{figure}

The differential $\chi$ flux per electron on target (EOT) produced in  a thick target, neglecting the transverse development of the electromagnetic shower, is given by:
\begin{equation}\label{eq:flux}
\begin{split}
\frac{d\Phi}{dE_\chi d\zeta_\chi d\phi_\chi} = \frac{1}{2\pi}\frac{N_{A}}{A_T}{\rho_T} \int dE_e \;\left(T_\pm(E_e) \frac{d\sigma_S(E_e)}{dE_\chi d\zeta_\chi} +\right.
\\
\left.+\,T_+(E_e)\frac{d\sigma_{R}(E_e)}{dE_\chi}\delta(\zeta_\chi-f(E_e,E_\chi,m_\chi))
\vphantom{\frac{d\sigma_S(E_e)}{dE_\chi d\zeta_\chi)}}\right) 
\end{split}
\end{equation}
where $A_T$ and $\rho_T$, are,  the target material atomic mass and mass density,  $N_A$ is Avogadro's number, $T_-(E_{e^-})$ ($T_+(E_{e^+})$) are the electron (positron) differential track-length distribution in the thick target per EOT, and $T_\pm(E)\equiv T_-(E)+T_+(E)$. Here, $\frac{d\sigma_S}{dE_\chi d
\zeta_\chi}$ is the differential cross section for $\chi$ production via $\Apr$-strahlung with respect to the $\chi$ energy $E_\chi$, polar angle $\zeta_\chi\equiv\cos(\theta_\chi)$ and azimuthal angle $\phi_\chi$, integrated over all the other final state kinematic degrees of freedom, while $\frac{d\sigma_R(E)}{dE_\chi}$ is the differential cross section for $\chi$ production via electron-positron resonant annihilation. The $\delta$ function accounts for the kinematic correlation between $E_\chi$ and $\zeta_\chi$ described by the function $f(E,E_\chi,m_\chi)$ resulting from the two-body nature of the resonance production. The effect of the transverse development of the electromagnetic shower  further broadens the angular distribution of the $\chi$ flux. However, for the multi-GeV energy range considered in this work, this effect is small compared to the intrinsic angular distribution of the production cross section. For illustration, Fig.~\ref{fig:ChiSpectrum} shows the differential $\chi$ flux for  fermionic LDM with $m_\chi=10$~MeV. LDM particles from $e^+e^-$ annihilation populate the high-density band in the range $E_\chi=$0.1 -- 0.8~GeV, clearly showing a strong $E_\chi$ vs $\zeta_\chi$ correlation. 
Finally, we observe that the integrated $\chi$ flux scales as $\varepsilon^2$, and it is almost independent of $\alpha_D$.

Once produced, the $\chi$ particles propagate through the beam dump and the other materials (concrete shielding, dirt) between the dump and BDX-MINI. LDM crossing the BDX-MINI ECal can interact with the PbWO$_4$ atomic electrons via elastic scattering, $\chi e^- \rightarrow \chi e^-$, resulting in a recoiling $e^-$. The differential spectrum of scattered electrons with respect to the recoil energy $E_r$ is given by:
\begin{equation}
\begin{split}
    \frac{dN}{dE_r} = N_{A}\rho_{E} \kappa_E \int_\Omega dE_\chi d\zeta_\chi d\phi_\chi \frac{d\sigma_E(E_\chi)}{dE_r}\\
    \cdot \frac{d\Phi_\chi}{dE_\chi d\zeta_\chi d\phi_\chi}\cdot L(\zeta_\chi,\phi_\chi)\cdot \eta \; \; ,
\end{split}
\end{equation}
where $\rho_E=8.28$ g/cm$^3$ is the PbWO$_4$ mass density, $\kappa_E=0.413$ is the PbWO$_4$ average $Z/A$ ratio, $\frac{d\sigma_E}{dE_r}$ is the $\chi-e^-$ elastic scattering cross section, $L$ is the $\chi$ particles path length in PbWO$_4$, and $\eta$ is the detection efficiency. The integral is performed over the phase-space element $\Omega$ defining the BDX-MINI ECal fiducial volume. Since the elastic scattering cross section is proportional to $\alpha_D\cdot \varepsilon^2$, the overall normalization of the signal yield scales as $\alpha_D \cdot \varepsilon^4$.

\subsection{Simulation framework}

The differential electron and positron track-length distributions with respect to the lepton energy and angle ($T_-$ and  $T_+$ in Eq.\,\ref{eq:flux}) were computed using the \texttt{FLUKA} simulation package, version \texttt{2021.2.1}~\cite{Bohlen:2014buj,Ferrari:2005zk}. We adopted the official Hall A beam dump description, including geometry and materials, provided to us by the JLab Radiation Control Group~\cite{Kharas}. A custom \texttt{mgdraw} routine was used to score, for each electron/positron step in the thick target, the energy $E_e$ and the angle $\theta_e$ with respect to the primary beam axis, with a scoring weight equal to the step length. 

%For a given value of the LDM mass $m_\chi$, we employed a custom version of the \texttt{MadGraph4} framework to compute the LDM flux from the beam dump, properly modified to handle fixed-target processes\textcolor{red}{citazione}. 
In order to compute the LDM flux produced in  the beam dump, we used a custom version of the \texttt{MadGraph4} toolkit, properly modified to handle fixed-target processes~\cite{Alwall:2007st}. The nuclear form factor for the radiative $\Apr$ emission adopted the parameterization reported in Ref.~\cite{Bjorken:2009mm}. 
An independent simulation was performed for each of the LDM mass points considered in this work, with fixed coupling constant $\varepsilon_0=3.87\cdot10^{-4}$ (see Eq.\,\ref{eq:lagrangian}). To account for the energy distribution of electrons and positrons in the beam dump, each simulation consisted of multiple runs performed by varying the primary energy $E$ between 0 and $E_0$, where $E_0$ is the beam energy. The generated events from all runs were recombined together, with relative weights given by $T_\pm(E)$ (radiative emission) / $T_+(E)$ (resonant production). The electromagnetic shower angular spread was also accounted for by further rotating all events in the transverse plane by a random angle extracted from the full $T(E,\theta)$ distribution. The outcome of the procedure was a set of LDM fluxes, one for each mass point considered, normalized to EOT, for the fixed coupling value $\varepsilon_0$.

The LDM differential fluxes were used as input for the simulation of $\chi-e^-$ interactions in BDX-MINI. We used a GENIE-based Monte Carlo code to sample the LDM flux from the beam dump, propagate the particles to the BDX-MINI detector, and simulate the elastic scattering with atomic electrons~\cite{Andreopoulos:2009rq,Andreopoulos:2015wxa}. Specifically, we employed the GENIE Boosted Dark Matter Module~\cite{Berger:2018urf}, tuning the parameters to reproduce the phenomenology of the dark sector Lagrangian reported in Eq.~\ref{eq:lagrangian}, consistent with a SM-LDM coupling purely proportional to the electric charge, with no left/right chiral asymmetries. We developed a custom driver for LDM flux sampling, importing the BDX-MINI detector geometry through a GDML file exported from the corresponding GEANT4 implementation. This approach guarantees that no mismatches are present between the $\chi-e^-$ scattering and the subsequent detector response simulation. The output from GENIE was a set of unweighted events for the process $\chi e^- \rightarrow \chi e^-$, including the final state particle four-vectors, the interaction vertex, and the corresponding total number of EOTs. The detector response to these events was computed by processing them through the GEANT4-based simulation code of the BDX-MINI detector package, and then running the JANA-based reconstruction on the results (see Ref.~\cite{Battaglieri:2020lds} for further details). Finally, the detection efficiency was obtained by imposing the same analysis cuts adopted in the real data analysis to the reconstructed Monte Carlo observables, in particular the veto response and total energy deposition in the ECal.

\section{\label{sec:analysis} Data analysis}
\subsection{\label{sec:data_reduction} Data reduction}

BDX-MINI operated parasitically to the experimental program in Hall A, accumulating approximately six months of data in 2020, corresponding to 2.56$\cdot$10$^{21}$ EOT. During most of the run, the Hall A beam dump received 2.176\,GeV electron beam with currents ranging from a few up to 150 $\mu$A. Cosmic-ray data were collected during the same period when the accelerator was down, either during maintenance days, or when the RF cavities tripped off for a long enough time. The beam-off time accounted for about $\sim$50$\%$ of the data-taking period. In addition, at the beginning of the run, Hall A experiment received a 10.381\,GeV energy beam with currents ranging from 5 to 35 $\rm \mu A$. This high energy run produced muons with sufficient energy to penetrate the shielding surrounding the dump and reach the detector with horizontal muons that were used for calibration. 

The offline data reconstruction procedure is described in detail in Ref.~\cite{Battaglieri:2020lds}.
%The off-line data reconstruction, implemented in the "JLAB Data Analysis Framework" (JANA).
%In addition to this procedure, an algorithm to perform a pulse shape analysis based on the cross-correlation technique was developed to reject electronic noise. An example of noise and signal waveforms is shown in Fig.~\ref{}.  
For each event, the recorded waveforms were processed to extract the corresponding charge, time, and amplitude.
For each PbWO$_{4}$ crystal, the integrated charge was converted into energy by applying proper calibration constants determined with muons generated in the beam dump during the high energy run.
The calorimeter calibration constants  were stable within 10\% over the entire measurement period~\cite{Battaglieri:2020lds}. 
In the data analysis we imposed a minimum energy threshold of 6~MeV to each crystal. 
%Since the data-taking lasted for several months, the calibration stability for the entire measurements period was studied using a procedure based on cosmic-rays described in Ref\cite{Battaglieri:2020lds}. The result, shown in Fig~\ref{}, indicates that the system was stable to within 10\,$\%$ over the data-taking period. \\
\begin{figure}[b]
    \centering
    \includegraphics[width=.48\textwidth]{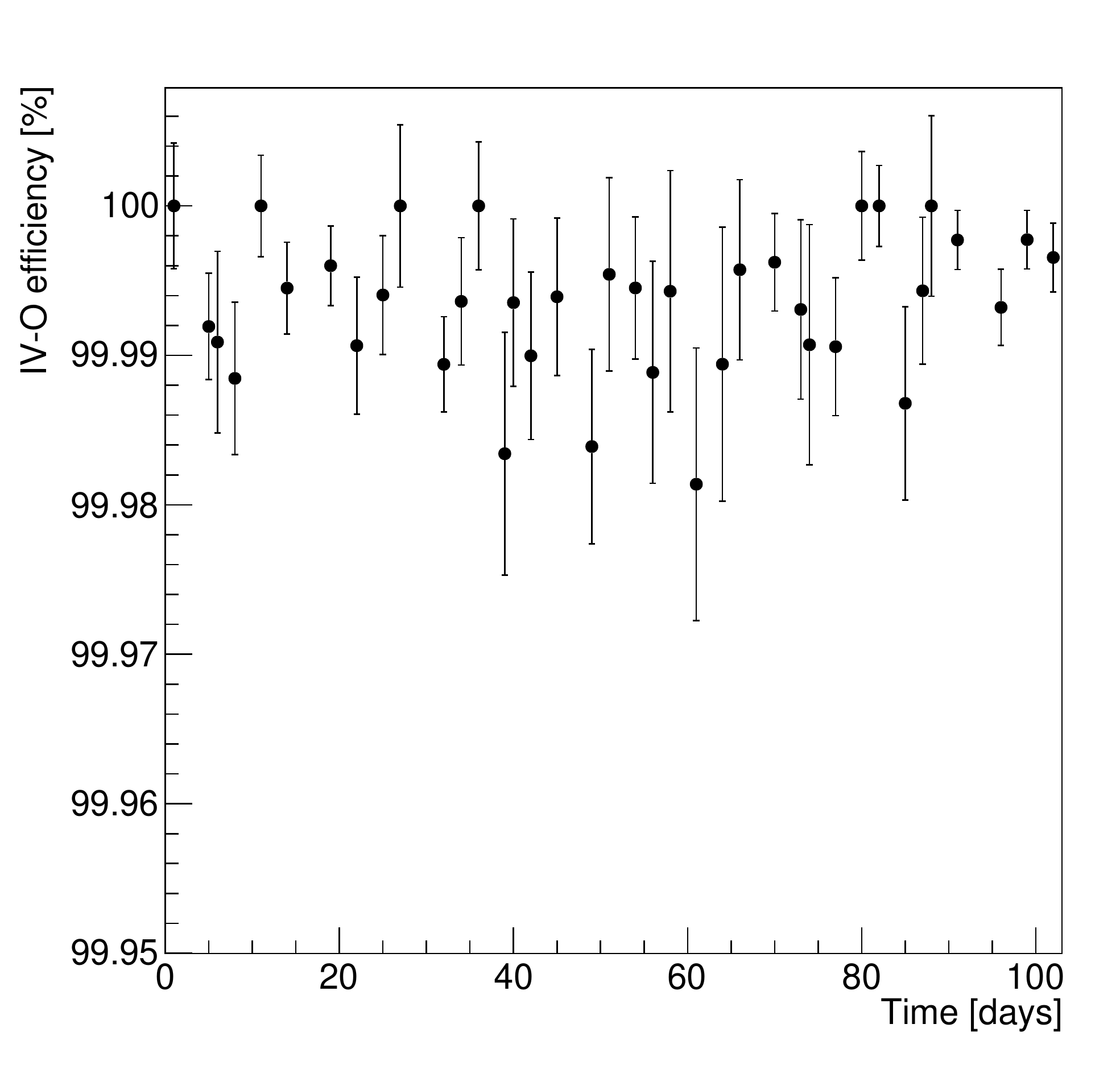}
    \caption{IV-O detection efficiency stability over the entire measurement period.}
    \label{fig:IVefficiency}
\end{figure} 

For plastic scintillator detectors, we used the amplitude of each signal normalized to its single photo-electron (p.e.) value and its corresponding time. As shown in Ref.~\cite{Battaglieri:2020lds}, the two veto systems are characterized by a detection efficiency compatible with 100\%. The stability of each veto response over time was studied using cosmic-ray data.  The response in the vertical sides of the IV  (OV) was measured by selecting hits in the sides of the OV (IV) and the upper caps to select cosmic muons with well defined trajectories. The response in the caps was determined by selecting perpendicular cosmic-ray tracks with a significant release of energy in caps ($>$~5.0 .p.e.) other than the one under study and no activity in the side veto channels. The selection procedure was repeated for each run indicating that each system was stable to better than 0.1$\%$ over the data-taking period. An example of the stability response as a function of time for IV sides (IV-Octagon) is shown in Fig. \Ref{fig:IVefficiency}.    

%{\it 1) data reduction (data reconstruction Sec.4.1( wave form analysis), Calibration from EPJA Sec.3, Stability (Ecal from EPJ, Veto), Systematics (energy, position, rotation, veto threshold)),}

%\begin{itemize}
%    \item Summary of EPJ
%    \begin{itemize}
        %\item Hints at noise filtering
        %\begin{itemize}
        %    \item Electrical noise in the board filtered with an algorithm trained on waveforms classified by eye
        %\end{itemize}
%        \item Data reconstruction (see EPJ)
%        \begin{itemize}
%            \item \textbf{(EPJ)} Waveform reconstruction: waveform analysis, energy and charge measurement
 %           \item Variables reconstructed from data 
 %       \end{itemize}
    
 %   \item Calibrations (see EPJ)
%        \begin{itemize}
 %           \item \textbf{(EPJ)} ECal calibration (beam muons): hints at method used, results
 %       \end{itemize}
%    \item Stability 
  %      \begin{itemize}
%            \item \textbf{(EPJ)} ECal stability (cosmic muons): hints at method used for the study (cosmic muons used) and results (crystals stable within 10$\%$
  %      \end{itemize}
%   \end{itemize}
 %\item Add: veto stability studies (vertical muons) 
  %          \begin{itemize} 
                %\item Events used to study %veto response stability: vertical muons and triggers used 
                %\item Measurement of veto response (relative efficiency) 
          %      \item Results (system stable within less than 1$\%$
   %         \end{itemize}
%\end{itemize}

\subsection{\label{sec:data_samples} Data Samples}
All  collected data were analyzed in  the very same way and then divided into two samples: beam-on, which corresponded to more than 10~$\mu$A requirement on current in Hall A, and beam-off. The beam-off sample was used to determine the cosmic-ray background in the beam-on sample after normalizing by the corresponding time. We emphasize that beam-on and beam-off data  were collected at almost the same time, so any  detector drift  would track together. 

\begin{figure}[t]
    \centering
    \includegraphics[width=.49\textwidth]{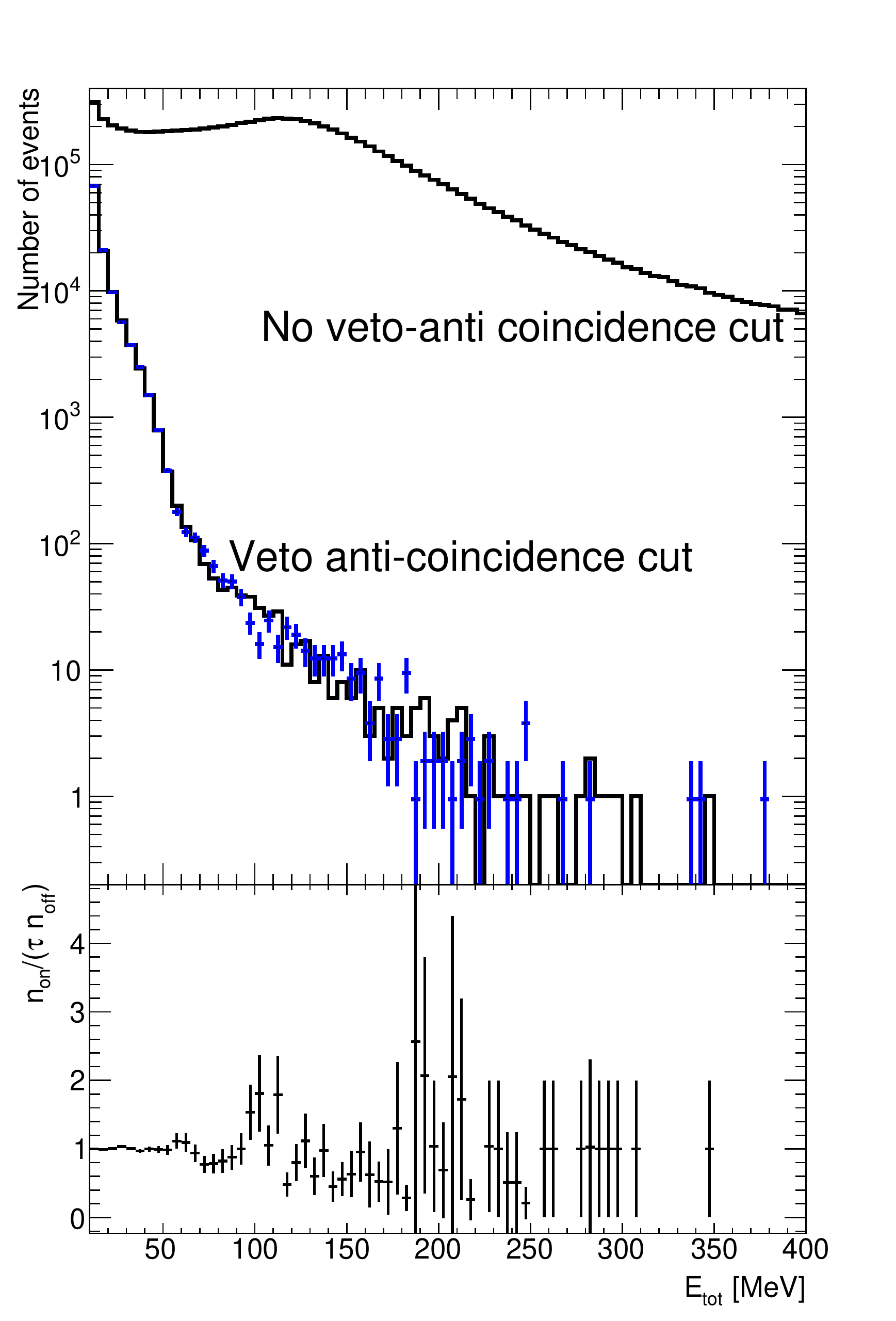}
    \caption{Top panel: in black, the ECal total energy
    distribution for the beam-on data sample, with and without the veto system's  anti-coincidence cut. In blue, the beam-off anti-coincidence data sample, properly normalized to take into account the difference in running time.
    Bottom panel: the ratio between the beam-on and the scaled beam-off anti-coincidence distributions.
    }
    \label{fig:DataReduction}
\end{figure}

The data reduction is illustrated in Fig.\,\ref{fig:DataReduction}, where we show the total energy distribution in the ECal before and after requiring the anti-coincidence with vetoes. A rejection factor between 2 and 4 orders of magnitude was found, depending on the amount of energy collected in the calorimeter. The plot also reports the ratio between beam-on and beam-off distributions, appropriately normalized,  showing that there is no statistical difference between the two data sets.

\subsection{\label{sec:background} Backgrounds}
Cosmic rays and beam induced neutrinos represent the two source of backgrounds in the BDX-MINI experiment. Neutrinos have a much lower interaction rate in the detector compared to the dominant cosmogenic contribution, but the neutrino background is irreducible. In this section we report on the study of cosmic backgrounds based on the beam-off data sample and summarize the neutrino background estimates based on Monte Carlo simulations.  

\subsubsection{Cosmic background}
 The cosmic background was determined studying data collected  during beam-off time. As described in Ref.~\cite{Battaglieri:2020lds}, each BDX-MINI veto is characterized by a high cosmic-ray rejection efficiency that increases by combining the information from both inner and outer vetoes. In fact the average cosmic rate for an ECal energy threshold of 40~MeV was $\sim$1.9~Hz, suppressed by 
 %a factor $\sim 5\times10^4$ 
 three to four orders of magnitude when we required no activity in both vetoes. In the data analysis, events were discarded if the signal of one SiPM of a veto system exceeded a threshold of 5.5 p.e., or at least two SiPMs of the same veto measured a signal above 2.5~p.e within a time window of 200~ns. Figure \ref{fig:EnergySpectrum} shows the ECal reconstructed energy distribution for events in anti-coincidence with both veto systems (green dots).
 
To demonstrate that cosmic events measured during beam-off are representative of beam unrelated background in the beam-on data set, we developed an ad-hoc procedure based on the event topology to identify vertical cosmic muons from a possible LDM hit (vertical down-warding vs.~beam dump originated horizontal tracks).
Vertical (cosmic muon) tracks were selected by requiring a significant release of energy ($>$6~MeV) in at least one crystal in the ECal top part  and in at least one crystal in the bottom part. In addition, we required a signal above 6 p.e. in all veto caps and no activity in the vertical veto sections. 
Since each BDX run included a combination of beam-on and beam-off time, we were able to compare  the beam-on/off vertical cosmic muon rates per each  run. 
\begin{figure}[htp]
    \centering
    \includegraphics[width=.51\textwidth]{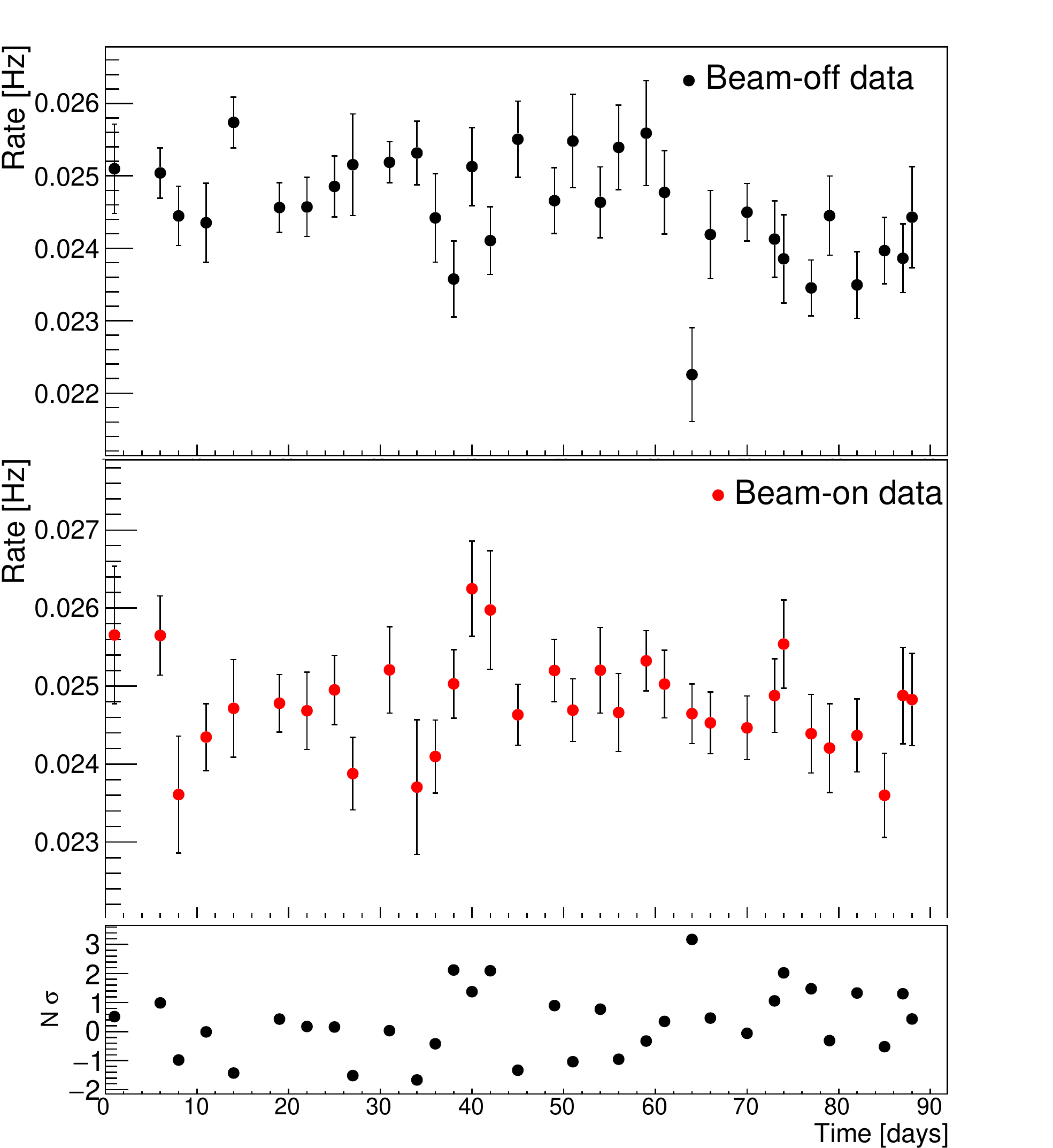}
    \caption{Vertical cosmic muon rate as function of time measured during beam-off (top panel) and beam-on data taking (middle panel). The comparison between beam-on and beam-off rate in terms of standard deviations is shown in the bottom panel.}
    \label{fig:VerticalMuon}
\end{figure} 
Figure \ref{fig:VerticalMuon} shows the vertical cosmic muon rate as a function of  run number for beam-off (top panel) and beam-on (middle panel). The results indicate that, although we cannot exclude a long-term variation within the same run, beam-on  and beam-off rates  are compatible within 2$\sigma$ as shown in Fig.\ref{fig:VerticalMuon}-bottom panel. This comparison demonstrates that any possible long-term variations of cosmic rays affect the beam-on and beam-off data in the same way. Therefore, we can use the cosmic measurement performed during  beam-off time to estimate the beam unrelated background in the beam-on data set. 

\subsubsection{Neutrino background}
To estimate the neutrino background  we performed  Monte Carlo simulations.
In particular to simulate the production, propagation, and detector interaction of neutrinos produced by the interaction of the primary electron beam  with the Hall A beam dump, we used the procedure described in Ref.~\cite{Battaglieri:2020lds}. Neutrino production was determined using  FLUKA simulations with the same  setup adopted for the LDM production. Neutrino interactions with the detector and the surrounding material was simulated with the GENIE package v.3.0.6~\cite{Andreopoulos:2009rq,Andreopoulos:2015wxa}. 
GENIE output was used as an input to the BDX-MINI detector  GEANT4-based simulation code of the BDX-MINI. Finally, simulations were  processed with the same JANA-based reconstruction code (see Sec.\ref{sec:analysis}) used to process experimental data. The total number of events requiring no activity in both vetoes, and an energy threshold of 40 MeV on ECal was 5.8x10$^{-23}$ per EOT~\footnote{In reference~\cite{Battaglieri:2020lds}, a neutrino event yield of 1.1x$10^{-21}$/EOT was erroneously reported for an energy threshold of 200 MeV, after anti-coincidence cut. This error was due to a wrong marginalization of the neutrino distribution obtained from GENIE.} This corresponds to less than 1 neutrino background event detected in the the entire BDX-MINI run.% that was accumulated during the BDX-MINI measurement. 

%\begin{enumerate}
%    \item Cosmic background
%    \begin{itemize}
%        \item Beam off cosmic background 
%        \begin{itemize}
%            \item \textbf{(EPJ)} Anti coincidence with the veto to reject cosmic background: veto rejection capabilities (high suppression)
%            \item Study of the rate of anti-coincidence events 
 %           \item Problem of fluctuations (fluctuations $\rightarrow$ need comparison with beam on)
  %      \end{itemize}
  %      \item Beam on cosmic background
  %      \begin{itemize}
  %          \item Beam-on background evaluated as the beam-off background scaled for the ratio of the times
   %     \end{itemize}
%        \item Validation (study of vertical muons)
%        \begin{itemize}
%            \item Ratio of beam-on and beam-off background compared between beam-on and beam-off data using vertical muons (no bias)
 %       \end{itemize}
 %   \end{itemize}
 %   \item Neutrino background
 %   \begin{itemize}
  %      \item \textbf{(EPJ)} Simulations: flux generated with FLUKA, interaction with GENIE; same detector implementation as DM
  %      \item Results (neutrino background negligible with newest simulations)
  %  \end{itemize}
%\end{enumerate}

\subsection{\label{sec:statistical_approach} Statistical Approach}
In order to derive an upper limit in the LDM parameters space, we adopted the following statistical procedure, based on the so-called ``On/Off'' problem (also called ``Li\,\&\,Ma'')~\cite{Gillessen:2004pm}. We started by considering a simple counting model, and dividing the measured data in two categories, the ``beam-on'' and the ``beam-off'' ones, characterized by overall time durations $T_{\rm on}$ and $T_{\rm off}$, respectively. We denoted as $n_{\rm on}$ and $n_{\rm off}$ the total number of  \textit{measured} events in each category. We assumed that these observables are distributed according to Poisson statistics, with average values $\mu_{\rm on}$ and $\mu_{\rm off}$, respectively. During the beam-off time, all measured events are due, by definition, by the beam unrelated (cosmic ray) background, while during beam-on time, the contribution from beam-induced (neutrino) background and from the LDM signal should also be added. Therefore, $\mu_{\rm on}=\mu_{c}+\mu_{b}+S$, where $S$ is the expected signal event yield, $\mu_{\rm off}=\mu_{c} \cdot \tau$ with $\tau=T_{\rm off}/T_{\rm on}$, and $\mu_b$ is the expectation for the beam-related backgrounds.
The following likelihood expression can thus be used to describe the measured dataset:
\begin{equation}\label{eq:unbinned_likelihood}
    \mathcal{L}=P(n_{\rm on};\mu_{c}+\mu_{b}+S)\cdot P(n_{\rm off};\mu_{c}\cdot\tau) \; \;
\end{equation}
where $P(n;\mu)$ denotes a Poisson distribution with average value $\mu$, computed for a number of events equal to $n$.
This likelihood expression can be employed to derive a test statistics to be used for the upper limit determination by treating $\mu_{c}$ as a nuisance parameter, while $\mu_{b}$ is uniquely determined from Monte Carlo simulations. It should be observed that, in this model, the parameter-of-interest (POI) $S$ represents the average event yield within the signal region due to any source other than the background ones. 
Therefore, the obtained upper limit on this parameter would be, by definition, model independent, and could be used to test \textit{any} BSM model against the measured data.

\begin{figure}
    \centering
    \includegraphics[width=.48\textwidth]{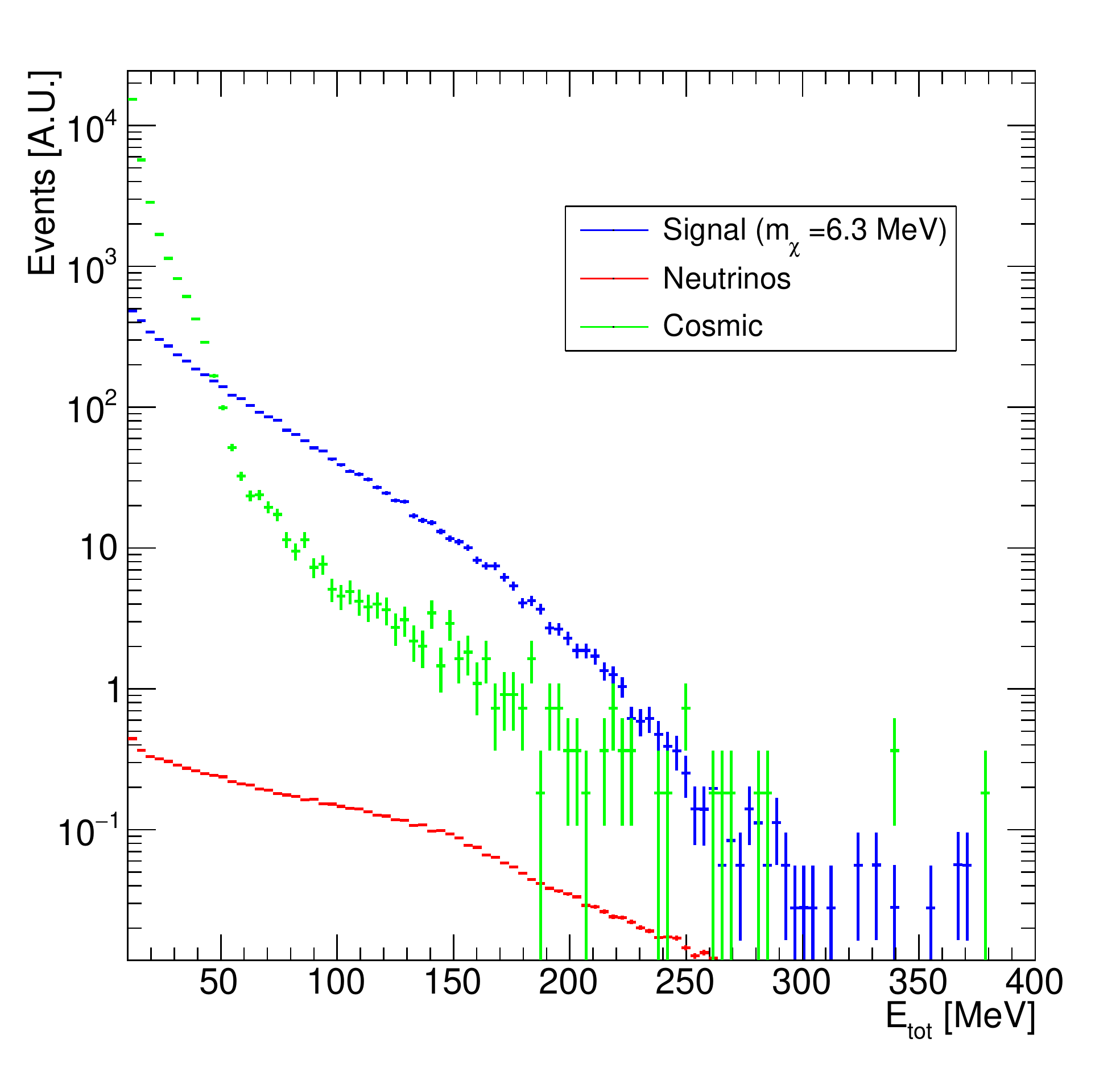}
    \caption{Deposited energy spectrum in the BDX-MINI ECal for anti-coincidence events associated to beam-related (red) and beam-unrelated (green) background, compared for the signal. The dark green points correspond to the total background yield. The 
    expected energy distribution for fermionic LDM with $m_\chi=6.3$ MeV was selected and arbitrarily normalized for illustration.
    \label{fig:EnergySpectrum}}
\end{figure}

We refined this model in order to include in the upper-limit evaluation procedure the measured energy deposition in the BDX-MINI detector for each event. We expect that, by increasing the observables employed in the statistical analysis, and thus the available information, the obtained upper limit would significantly improve, as already observed in similar studies~\cite{deRomeri:2020kno}. We expect the largest improvement to be in the $\Apr$ mass region where the $e^+e^-$ annihilation production mechanism is prominent, due to the peculiar energy deposition spectrum associated of signal events, which is significantly different than the background (see also Fig.~\ref{fig:EnergySpectrum}).
We further divided each set of data  into discrete bins of the total BDX-MINI measured energy $E_{\rm tot}$. By calling $n^j_{\rm on}$ and $n^j_{\rm off}$ the number of measured events in the $j^{\rm th}$ energy bin, the likelihood model now reads:
\begin{equation}
    \mathcal{L}=\prod_{j}\left[P(n^j_{\rm on};\mu^j_{c}+\mu^j_{b}+\alpha^j\cdot S)\cdot P(n^j_{\rm off};\mu^j_{c}\cdot\tau) \right]\; \;,
\end{equation}
where the $\{\mu^j_c\}$ coefficients are treated as nuisance parameters, and the $\{\mu^j_b\}$ parameters are obtained from the Monte Carlo simulation of the beam-induced background. As before, the POI $S$ is the average signal event yield. Finally, the coefficients $\{\alpha^j\}$ represent the fraction of signal events expected in the $j^{\rm th}$ energy bin (see Section\,\ref{sec:sensitivity_optimization} for their specific assignment). We computed these through a Monte Carlo simulation of the signal events, performed independently for each LDM model considered in this work and for each tested $\Apr$ mass value. We note that the coefficients $\{\alpha^j\}$ are not independent, but satisfy, by definition, the relation $\sum_j \alpha_j=1$.

To extract the upper limit, a one-sided profile-likelihood test statistics was used:
\begin{equation}
    q(S) =
    \begin{cases}
      -2 \log{\lambda(S)} & \text{if $S$ $\geq$ $\hat{S}$}\\
      0 & \text{if $S$ $<$ $\hat{S}$ \; \; ,}\\
    \end{cases}       
\end{equation}
where 
$\lambda(S)=\frac{\mathcal{L}(S,
\hat{\vphantom{\rule{1pt}{6.5pt}}\smash{\hat{\theta}}}
)}{\mathcal{L}(\hat{S},\hat{\theta})}$ 
is the profile likelihood, and $\theta$ denotes the ensemble of all nuisance parameters~\footnote{We used the ``traditional'' notation $\hat{a}$ to denote the value of a likelihood parameter at the absolute function maximum, and
$\hat{\vphantom{\rule{1pt}{5.5pt}}\smash{\hat{a}}}$ to denote the value of a parameter that maximizes $\mathcal{L}$ for a fixed value of the POI $S$.}.

To translate the upper limit on $S$ to an upper limit for the LDM coupling $\varepsilon$, we computed through Monte Carlo the signal event yield $S_0$ for a nominal coupling value $\varepsilon_0$ using the relation $S = S_0 \cdot (\varepsilon/\varepsilon_0)^4$. We underline that the upper limit on $S_0$ was computed independently for all LDM models and $\Apr$ mass values considered in this work, due to the dependency associated to the LDM energy deposition spectrum in the BDX-MINI detector affecting the coefficients $\{\alpha_j\}$.

To incorporate the systematic uncertainties associated with the measurement in the upper limit extraction procedure, we modified the likelihood model for the experiment as follows. 
The systematic uncertainty on the detector position and alignment, as well as that on the veto response, does not affect directly the obtained upper limit on the number of signal events $S$, rather the corresponding projection on the LDM parameters space. More precisely, these systematic uncertainties affect the evaluation of the nominal signal strength $S_0$, and thus the upper limit on $\varepsilon$. To evaluate this effect, we independently studied the dependency of $S_0$ on the vertical position of the detector in the beam pipe, $y$, on the detector rotation around the vertical axis, $\theta$, and on the scale of the veto calibration, $q$. For illustration, the signal yield variations corresponding to a $1\sigma$ modification of these parameters are summarized in Table~\ref{tab:systematics}.
To include these systematic uncertainties in the likelihood, we introduced the following parameterization:
\begin{equation}\label{eq:S0}
    S_0(y,th) = S_0^{\rm nom}\cdot A(y)\cdot B(th) \; \; \;,
\end{equation}
where $A(0)=1$, $B(1)=1$ correspond to the nominal scenario, giving the event yield $S_0^{\rm nom}$. The functions $A$ and $B$ were obtained from Monte Carlo simulations, changing the detector position and the veto thresholds $th$, and evaluating the expected signal yield, normalized to $S_0^{\rm nom}$. This study was repeated independently for both LDM models and for all the LDM masses considered in this work. We neglected the effect of the detector rotation, since it results to a much smaller change in $S_0$. In the likelihood, given the linear relation between $S_0$ and $S$, we multiplied the latter by $A(y)\cdot B(th)$, including two Gaussian PDF terms to constrain the nuisance parameters $y$ and $th$: 
\begin{equation}
    \mathcal{L} \rightarrow \mathcal{L}\cdot G(0;y,\sigma_y) 
    \cdot G(1;th,\sigma_{th})\; \;.
\end{equation}

\begin{table}[t!]
    \centering
    \begin{tabular}{l|l|l}
       \textbf{Syst. uncertainty term} & \textbf{1-$\sigma$ variation} & \textbf{$\Delta S_0 /S_0$}  \\
       \hline
        Detector vertical position  & 5 cm & 0.07 \\
        Detector rotation           & 5$^\circ$ & 0.025 \\
        Veto thresholds             & $\simeq 2$ phe & 0.05 \\
        \multirow{2}{*}{Energy calibration}         & \multirow{2}{*}{10$\%$} & $\chi$ mass-dependent \\
        & & maximum $\simeq 0.1$ 
    \end{tabular}
    \caption{ \label{tab:systematics}Effect of the detector configuration systematic uncertainties on the nominal LDM signal yield.}
\end{table}

\begin{figure}[t]
    \centering
    \includegraphics[width=.48\textwidth]{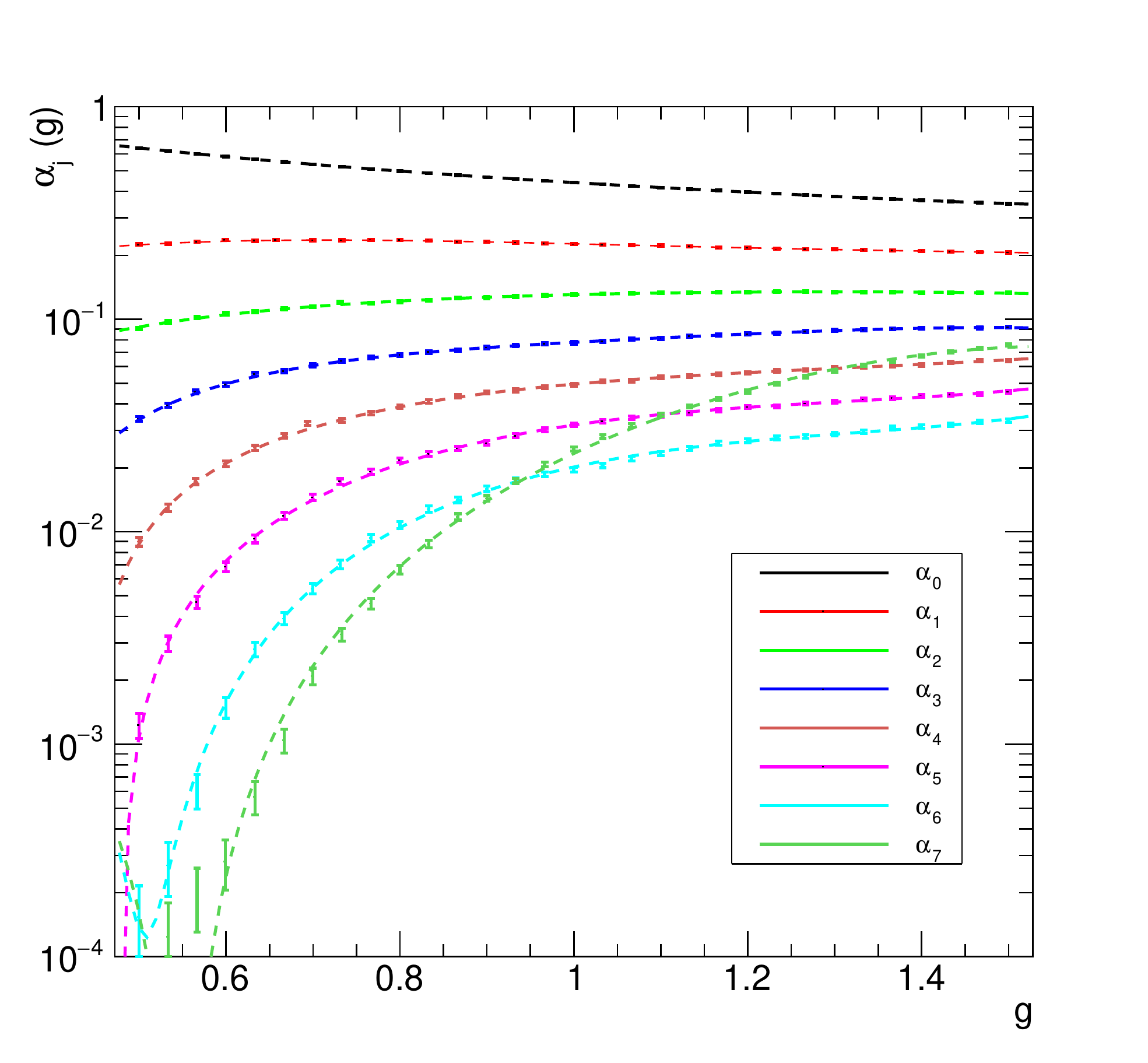}
    \caption{Plot of the functions $\alpha^j(g)$ for the illustrative case of fermionic LDM with $m_\chi=8.5$~MeV. The graphs show the results obtained from the Monte Carlo runs performed with different energy calibration scales, while the functions are the corresponding polynomial interpolations. The case $g=1$ corresponds to a perfect agreement between the real detector response and the one implemented in the Monte Carlo simulation.}
    \label{fig:relative_fraction_sys}
\end{figure}

\begin{figure}[t]
    \centering
    \includegraphics[width=.4\textwidth]{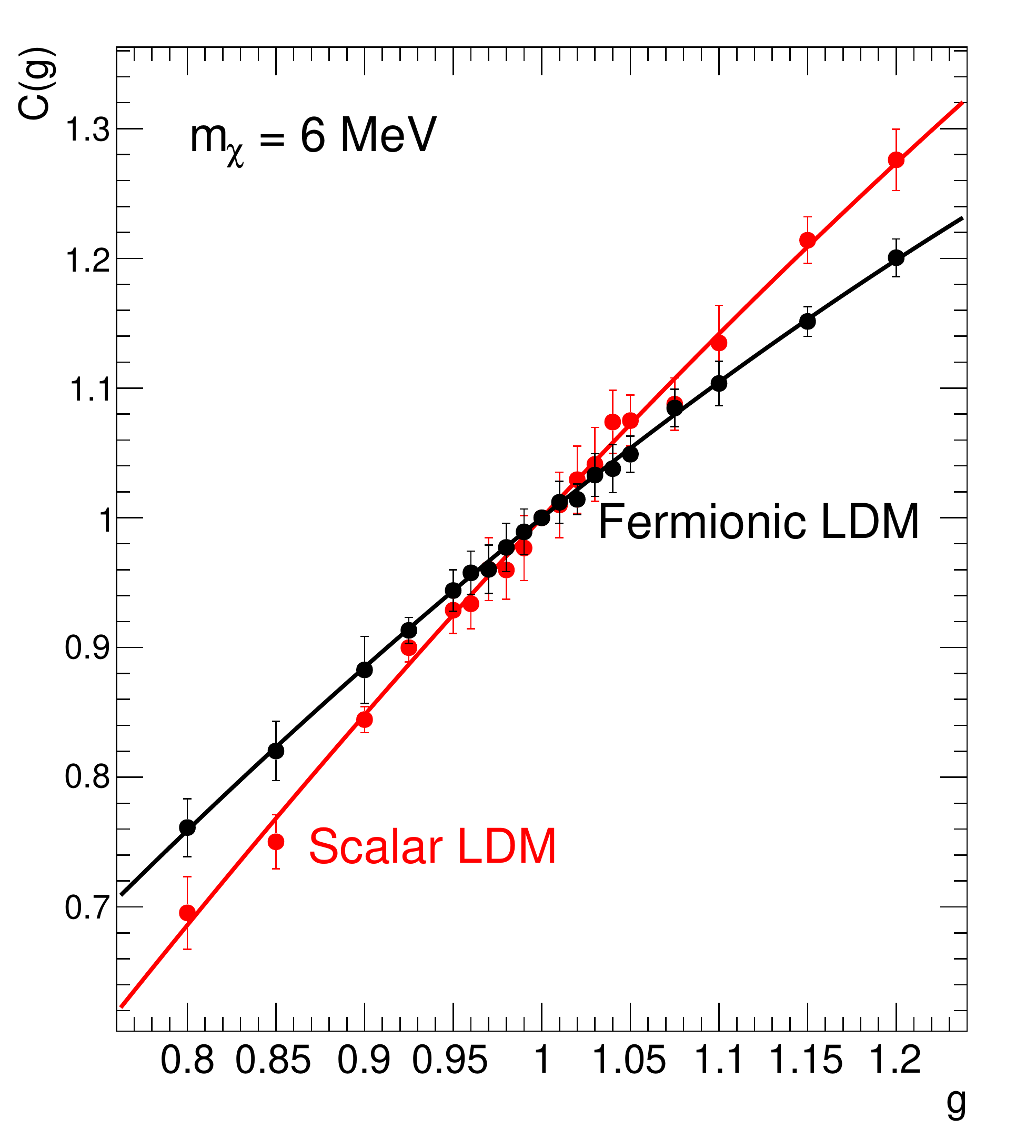}
    \caption{The function $C(g)$ describing the effects of a variation of the BDX-MINI ECal energy scale, computed for $m_\chi=6$~MeV, for the fermionic (black) and scalar (red) case. Each point is the result obtained processing Monte Carlo signal events through the full reconstruction chain, with a fixed variation of the energy scale. The two curves are the results of a best fit with a second order function, with the constraint $C(1)=1$.}
    \label{fig:Cg}
\end{figure}

In contrast, the uncertainty in the ECal energy scale affects both the nominal signal yield $S_0^{\rm nom}$ and the coefficients $\{\alpha^j\}$. The changes in the fractions $\{\alpha^j\}$ are due to the modification of the shape of the $E_{\rm tot}$ distribution due to variations of the energy response of each crystal. This results in energy bin-migration.  
This effect is primarily associated with systematic uncertainties that alter the energy scales of all crystals simultaneously, such as a small mismatch between the true PbWO$_4$ crystal density and the value implemented in the Monte Carlo, or a incorrect description of the passive materials surrounding the detector that were traversed by muons generated by the 10.38-GeV beam, whose ionization signal was used to infer the crystals energy calibration constant. Denoting with $g$ the ratio between the assumed energy calibration constants and the (unknown) real ones, we introduced the parameterization $\alpha^j \rightarrow \alpha_j(g)$, with the constraint $\alpha_j(1)=\alpha^j$. We computed the functions $\alpha_j(g)$ using Monte Carlo by performing simulations with different values of the calibration constants. Since each LDM model has its own specific energy distribution, we repeated the calculation for each $\Apr$ mass value, and for both the fermionic and the scalar case. As an example, Fig.~\ref{fig:relative_fraction_sys} shows the behavior of these functions for the fermionic LDM case, with $m_\chi=8.5$~MeV.  The nuisance parameter $g$ was constrained by adding a further Gaussian PDF term to the likelihood:
\begin{equation}
    \mathcal{L} \rightarrow \mathcal{L}\cdot G(1;g,\sigma_g) \; \;,
\end{equation}
where $\sigma_g=0.1$ is the uncertainty in the BDX-MINI \textit{absolute} energy scale. We observe that the functions $\alpha_j(g)$ still satisfy the constraint $\sum_j \alpha_j(g)=1$, since for each value of $g$ these have been computed by considering only the events within the $E_{\rm tot}$ signal region.
At the same time, a variation of the crystals energy scale would affect the nominal signal yield $S_0^{\rm nom}$ due to the lower energy threshold on $E_{\rm tot}$ that defines the signal region, as discussed in the next Section. We thus rewrote the expression of $S_0$ from Eq.~\ref{eq:S0} as:
\begin{equation}\label{eq:S1}
      S_0(y,th,g) = S_0^{\rm nom}\cdot A(y)\cdot B(th) \cdot C(g)\; \; \;,
\end{equation}
with the nominal condition $C(1)=1$. Similarly as before, we computed the function $C(g)$ from Monte Carlo. To account for the uncertainty on the \textit{relative} crystal-to-crystal response, also including the corresponding time fluctuations, we proceeded as follows. For each value of $g$, we performed different simulations, randomly extracting, for each crystal, a multiplicative energy correction factor from a Gaussian distribution with an average value $g$ and a standard deviation of $10\%$, which is the typical variation of the relative calibration constants observed during the full BDX-MINI run~\cite{Battaglieri:2020lds}. For each simulation, we computed the number of events within the signal region, normalized to $S_0^{\rm nom}$, finally quoting for $C(g)$ the average value of this distribution. For illustration, Fig.~\ref{fig:Cg} shows the $C(g)$ function for $m_\chi=6$~MeV.

The final expression of the likelihood used in the upper limit evaluation procedure reads:
\begin{align}\label{eq:binned_likelihood}
    \mathcal{L}=&\prod_{j}\left[P(n^j_{\rm on};\mu^j_{c}+\mu^j_{b}+\alpha^j(g)\cdot S\cdot A(y)\cdot B(th) \cdot C(g))\right. \cdot \nonumber\\
    &\left. \cdot P(n^j_{\rm off};\mu^j_{c}\cdot\tau) \right] \cdot\nonumber\\
    & \cdot G(0;y,\sigma_y)\cdot G(1;th,\sigma_{th})\cdot G(1;g,\sigma_g)
\end{align}

In this procedure, we neglected any correlation among the systematic uncertainty sources - for example, a modification in the detector position could change the energy spectrum of LDM particles passing through it. The validity of this approximation was explicitly checked by looking at shapes of the functions $\alpha_j(k)$ for different detector configurations. No significant variations with respect to the nominal case were found.

We implemented this statistical model with the \texttt{RooStats} software~\cite{Moneta:2010pm}, using a toy-MC approach to determine the PDF of the $q(S)$ test statistics, and extract from this the $90\%$ CL limit on the POI $S$. In the next section, we describe the procedure adopted to fix the analysis parameters and the likelihood model (number of energy bins, thresholds), reporting  the obtained results.

\section{\label{sec:results} Results}

\subsection{\label{sec:sensitivity_optimization} Sensitivity optimization}
The reach of the experiment was optimized considering simultaneously the effect 
of data analysis cuts on background minimization and signal maximization.

\begin{figure}
    \centering
    \includegraphics[width=.48\textwidth]{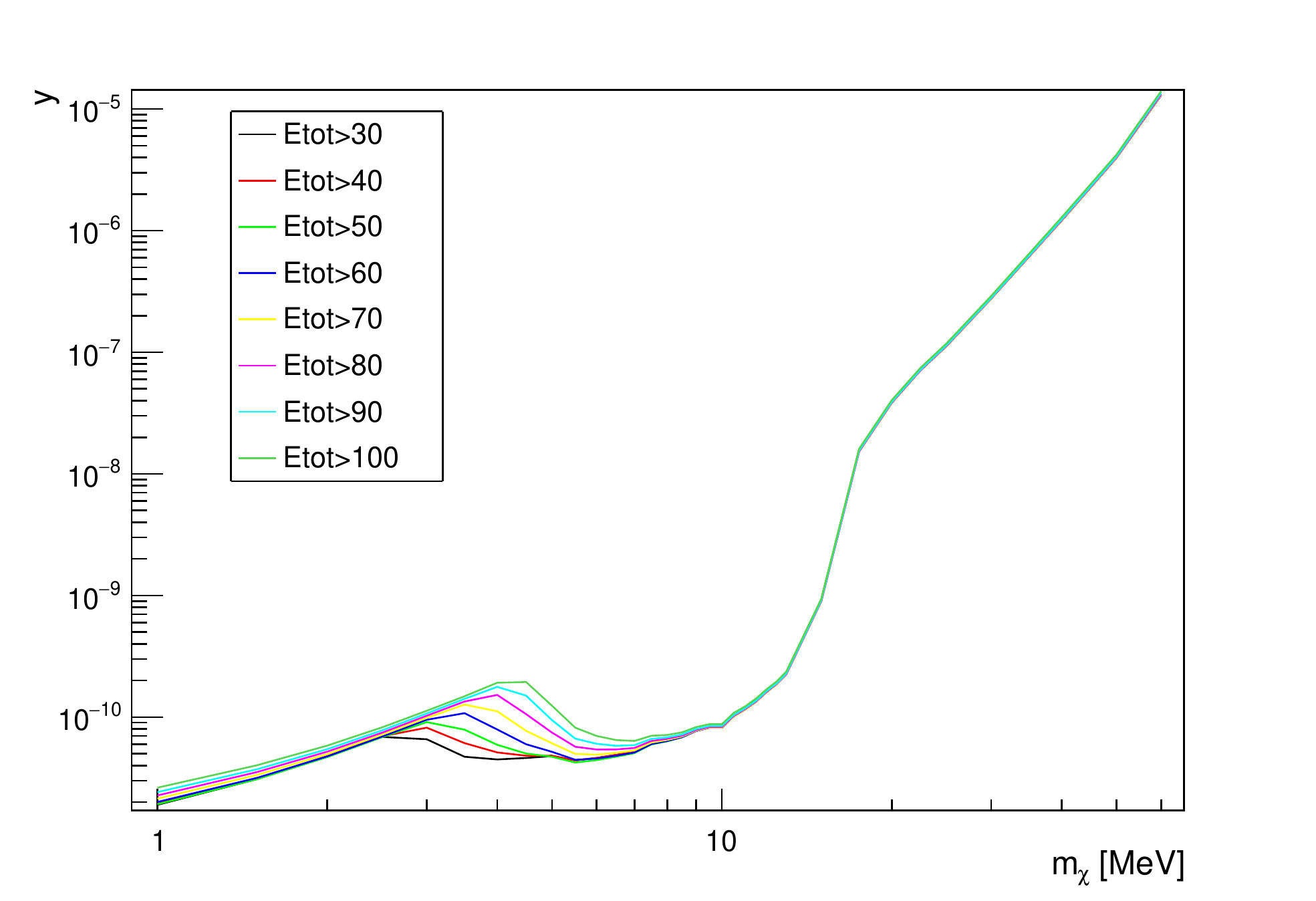}
    \caption{Expected sensitivity for different energy. The average exclusion limit is obtained using the procedure presented in text. Different colors correspond for different cuts on the minimum energy. The same bin width and energy range was used for all the cuts (the last bin size was set to include events with energy up to 600 MeV). The sensitivity is strongly dependent on the energy cut for $\chi$ masses of few MeV.
    \label{fig:compare_E_cut}}
\end{figure}

Beam-off data were assumed to correspond to the cosmogenic component of beam-on data. We only used events accepted by the anti-coincidence condition reported in \ref{sec:background} to reduce the cosmic background. We used the procedure described in Sec.~\ref{sec:statistical_approach} to evaluate the average upper limit on the number of signal events under the hypothesis $S=0$, also referred to as \textit{sensitivity}. Through MC simulations, we converted this to the sensitivity on the parameter $y$ defined in Sec.~\ref{sec:intro}, for different values of $m_\chi$. Specifically, we evaluated how the sensitivity varies using different selection cuts: if a cut suppresses the background, while preserving the signal, the exclusion limit is expected to be more stringent.
As a first step, we optimized the sensitivity by varying the number and the size of the energy bins. We found that the sensitivity is mostly dependent on the choice of the minimum total energy, and while it depends weakly on the other analysis parameters. Fig.~\ref{fig:compare_E_cut} shows the expected sensitivity using the same bin width but different minimum energy. The best sensitivity was achieved by dividing the ECal energy spectrum  into seven 45 MeV-wide bins from $E_{tot}=40$ MeV to $E_{tot}=355$ plus an 8th bin from $E_{tot}=355$ to $E_{tot}=600$ MeV.

The sensitivity obtained using Eq.~\ref{eq:binned_likelihood} was compared to the one obtained without any binning (Eq. \ref{eq:unbinned_likelihood}) finding an improvement in the exclusion limit of about a factor of two  over the entire mass range.

To further enhance the reach we studied the sensitivity as a function of  other measured quantities expected to be different for signal and background: hit multiplicity (number of crystals with an energy deposition over a channel dependent threshold), electromagnetic shower direction (defined as a fit of the position of different energy depositions), and energy distribution (defined as the fraction of energy deposited outside the most energetic hit). Although all these variables had some discrimination power, none provided  a significant sensitivity improvement. For sake of simplicity, we then decided to quote 
an exclusion plot only based on the total energy binning.

This study was performed in two scenarios:  scalar and fermionic LDM. The results were similar, achieving  the best sensitivity when events with total energy $>40$ MeV and satisfying  the veto anti-coincidence condition were considered. The data have then been divided into eight bins according to the total energy deposited: the energy range between $40$ and $355$ MeV was split into seven equally spaced 45 MeV-wide bins, and a single bin was used for all events from  $355$ to $600$ MeV.

\subsection{\label{sec:upper_limits} Upper Limits}
\begin{figure}[t]
    \centering
    \includegraphics[width=.48\textwidth]{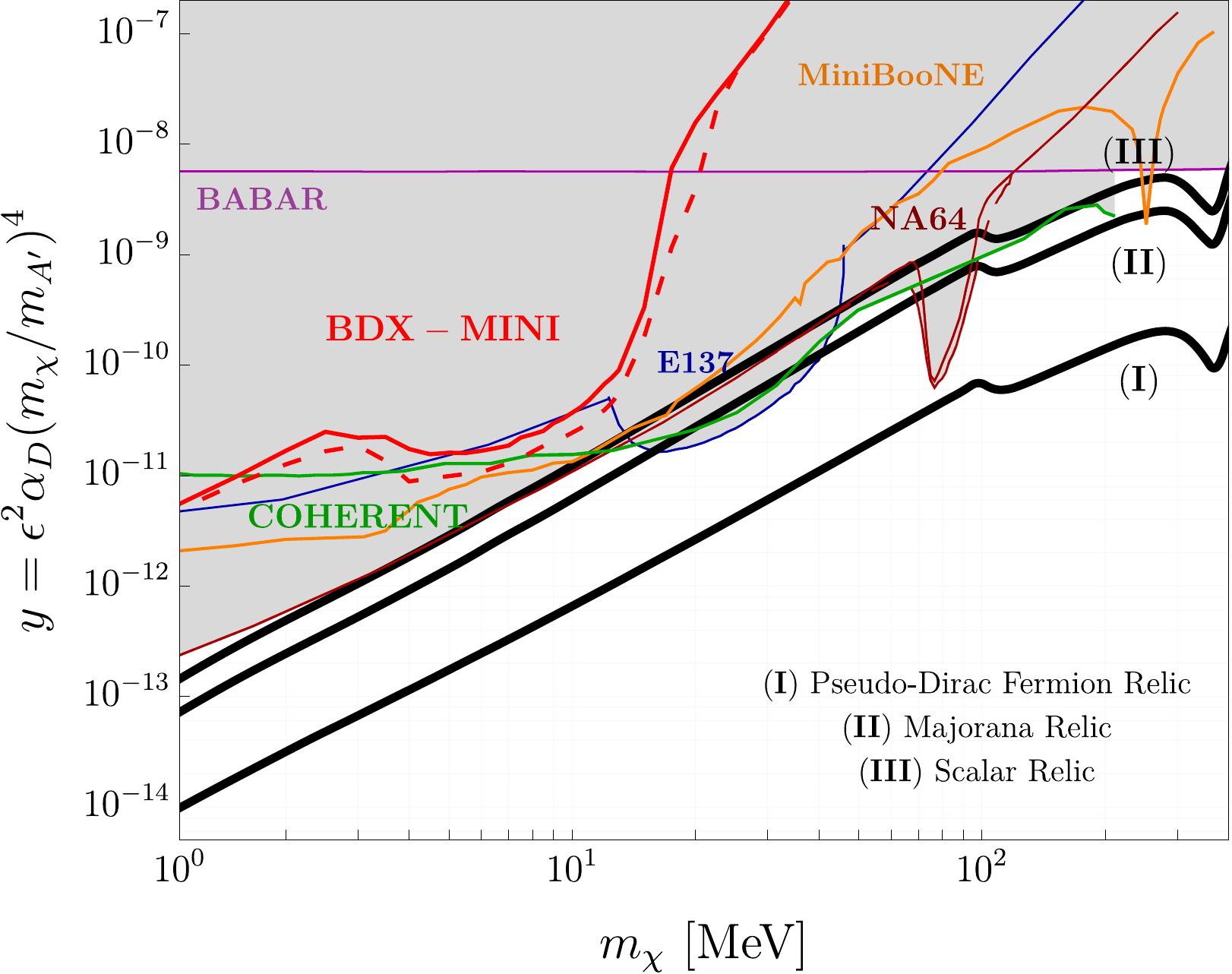}
    \caption{BDX-MINI exclusion limits assuming $\alpha_D = 0.1$ and $m_{A'}=3m_\chi$ are shown as red lines: the continuous line represents the exclusion limit for scalar LDM and the dashed line for fermionic LDM. The thick black lines represent the relic target. The other colored lines show the exclusion limits from BABAR~\cite{BaBar:2017tiz}, NA64~\cite{Banerjee:2019pds,Andreev:2021fzd}, MiniBooNE~\cite{MiniBooNEDM:2018cxm}, E137~\cite{Batell:2014mga,Marsicano:2018glj}, and Coherent~\cite{COHERENT:2021pvd}. 
    \label{fig:reach}}
\end{figure}

In the last step of the analysis, we applied the procedure optimized on the beam-off data set only to the whole beam-on data set, which corresponds to an accumulated charge of $2.56 \times 10^{21}$ EOT. We refer to this process as ``unblinding.''
To avoid possible long-term fluctuations between beam-off and beam-on data samples,
only runs with a significant beam-on and beam-off time were included in the analysis (see Sec.~\ref{sec:background} for details). Starting from the measured yields of beam-on $n_{\mathrm{on}}=3623$ and beam-off events $n_{\mathrm{off}}=3822$ ($\tau=1.054$) we derived a $90 \%$ exclusion limit on the LDM yield $S$ through the statistical procedure described in Sec.~\ref{sec:statistical_approach}. MC simulations were used to translate this result to an exclusion limit on the LDM parameter $y$ for both fermion and scalar LDM. Fig. \ref{fig:reach} shows exclusion limits obtained from BDX-MINI data in the two models. 

As anticipated, we observe that resonant annihilation enhances the exclusion limits for $m_\chi$ in the ($3$ MeV $< m_\chi < 20$ MeV) range. In this range, the lowest limit on $y \sim 10^{-11}$ is obtained for $m_\chi = 4.5$ MeV for fermionic LDM, while for scalar LDM lower limits are obtained for $m_\chi = 4 $ MeV. The cut-off at low mass for resonant annihilation is determined by the energy detection threshold. The non resonant contribution is slightly less sensitive, but extends the reach to lower masses down to few MeV. In some selected kinematics, the exclusion limits set by BDX-MINI are comparable to the best existing upper limits.

\section{\label{sec:summary} Summary and outlook}
In this paper we report on the exclusion limits placed on the production of LDM by the pilot BDX-MINI experiment.
%In this paper, we reported the BDX-MINI results,  a beam-dump experiment optimized for Light Dark Matter search. 
The experiment ran in parallel to the Jefferson Lab Hall A physics program, accumulating the considerable charge of $2.56 \times 10^{21}$ EOT. We installed a limited-volume $\rm PbWO_4$ EM calorimeter, surrounded by a hermetic veto system, into an existing well downstream of the Hall A beam dump. This detector was exposed to any DM particles produced by a high intensity 2.176 GeV electron beam in the dump and penetrating the material between the dump and the detector. Given the limited beam energy, no Standard Model particles produced in the dump were expected to reach the detector, except for neutrinos. The main background source for the experiment was cosmic rays, carefully studied with data collected during the beam-off times that were interspersed with the beam-on periods.
Simulations of fermionic and scalar dark matter produced in the dump and interacting with electrons in the  BDX-MINI calorimeter were used to determine the sensitivity of the experiment.
The  analysis procedure to detect an excess of counts in the beam-on data sample was optimized using the fully blinded beam-off data sample.  A sophisticated statistical procedure, based on a maximum-likelihood approach, was applied to the deposited energy spectrum to optimize the experiment sensitivity. Sources of systematic errors, such as the energy calibration and the detector misalignment, were included in the analysis to provide a reliable and robust result.
Two possible scenarios, fermionic and scalar dark matter were investigated. 
%From the comparison of beam-on and beam-off yield, we observed that the number of recorded events during beam-on was statistically compatible with beam-off expectation. 
Based on the measured event yields during beam-on and beam-off periods, a 90$\%$ CL upper limit on LDM production was derived as a function of a hypothetical DM particle $\chi$ mass.
The sensitivity of this pilot experiment covers some kinematics comparable with flagship experiments such as NA64~\cite{Banerjee:2019pds,Andreev:2021fzd}. Considering the limited size of the BDX-MINI active volume this is a remarkable result, demonstrating the potential of the new generation of beam dump experiments in LDM searches.  The full BDX experiment has already been approved to run at Jefferson Lab and, with its improved sensitivity, it has great discovery potential to find hints of Dark Matter or, in case of a null result, to improve existing limits by up to two orders of magnitude.

\section*{Acknowledgments}
The authors would like to thank the JLab Directorate and Physics Division, and the Italian Istituto Nazionale di Fisica Nucleare for invaluable support during the entire project.
We would also like to thank the INFN technical staff for the excellent work in constructing the detector, Hall D Physics technical staff for careful installation, JLab Facilities for design, survey, and logistical support, JLab Networking/Computing for providing connectivity.
Special thanks to CLAS12 Collaboration and PANDA Collaboration  for providing the PbWO$_{4}$ crystals.
This material is based upon work supported by the U.S. Department of Energy, Office of Science, Office of Nuclear Physics under contract DE-AC05-06OR23177.

\bibliographystyle{apsrev4-1} % Tell bibtex which bibliography style to use
\bibliography{biblio}

%merlin.mbs apsrev4-1.bst 2010-07-25 4.21a (PWD, AO, DPC) hacked
%Control: key (0)
%Control: author (72) initials jnrlst
%Control: editor formatted (1) identically to author
%Control: production of article title (-1) disabled
%Control: page (0) single
%Control: year (1) truncated
%Control: production of eprint (0) enabled
\begin{thebibliography}{43}%
\makeatletter
\providecommand \@ifxundefined [1]{%
 \@ifx{#1\undefined}
}%
\providecommand \@ifnum [1]{%
 \ifnum #1\expandafter \@firstoftwo
 \else \expandafter \@secondoftwo
 \fi
}%
\providecommand \@ifx [1]{%
 \ifx #1\expandafter \@firstoftwo
 \else \expandafter \@secondoftwo
 \fi
}%
\providecommand \natexlab [1]{#1}%
\providecommand \enquote  [1]{``#1''}%
\providecommand \bibnamefont  [1]{#1}%
\providecommand \bibfnamefont [1]{#1}%
\providecommand \citenamefont [1]{#1}%
\providecommand \href@noop [0]{\@secondoftwo}%
\providecommand \href [0]{\begingroup \@sanitize@url \@href}%
\providecommand \@href[1]{\@@startlink{#1}\@@href}%
\providecommand \@@href[1]{\endgroup#1\@@endlink}%
\providecommand \@sanitize@url [0]{\catcode `\\12\catcode `\$12\catcode
  `\&12\catcode `\#12\catcode `\^12\catcode `\_12\catcode `\%12\relax}%
\providecommand \@@startlink[1]{}%
\providecommand \@@endlink[0]{}%
\providecommand \url  [0]{\begingroup\@sanitize@url \@url }%
\providecommand \@url [1]{\endgroup\@href {#1}{\urlprefix }}%
\providecommand \urlprefix  [0]{URL }%
\providecommand \Eprint [0]{\href }%
\providecommand \doibase [0]{http://dx.doi.org/}%
\providecommand \selectlanguage [0]{\@gobble}%
\providecommand \bibinfo  [0]{\@secondoftwo}%
\providecommand \bibfield  [0]{\@secondoftwo}%
\providecommand \translation [1]{[#1]}%
\providecommand \BibitemOpen [0]{}%
\providecommand \bibitemStop [0]{}%
\providecommand \bibitemNoStop [0]{.\EOS\space}%
\providecommand \EOS [0]{\spacefactor3000\relax}%
\providecommand \BibitemShut  [1]{\csname bibitem#1\endcsname}%
\let\auto@bib@innerbib\@empty
%</preamble>
\bibitem [{\citenamefont {Cebri\'an}(2022)}]{Cebrian:2022brv}%
  \BibitemOpen
  \bibfield  {author} {\bibinfo {author} {\bibfnamefont {S.}~\bibnamefont
  {Cebri\'an}},\ }in\ \href@noop {} {\emph {\bibinfo {booktitle} {{10th
  Symposium on Large TPCs for Low-Energy Rare Event Detection}}}}\ (\bibinfo
  {year} {2022})\ \Eprint {http://arxiv.org/abs/2205.06833} {arXiv:2205.06833
  [physics.ins-det]} \BibitemShut {NoStop}%
\bibitem [{\citenamefont {Arbey}\ and\ \citenamefont
  {Mahmoudi}(2021)}]{Arbey:2021gdg}%
  \BibitemOpen
  \bibfield  {author} {\bibinfo {author} {\bibfnamefont {A.}~\bibnamefont
  {Arbey}}\ and\ \bibinfo {author} {\bibfnamefont {F.}~\bibnamefont
  {Mahmoudi}},\ }\href {\doibase 10.1016/j.ppnp.2021.103865} {\bibfield
  {journal} {\bibinfo  {journal} {Prog. Part. Nucl. Phys.}\ }\textbf {\bibinfo
  {volume} {119}},\ \bibinfo {pages} {103865} (\bibinfo {year} {2021})},\
  \Eprint {http://arxiv.org/abs/2104.11488} {arXiv:2104.11488 [hep-ph]}
  \BibitemShut {NoStop}%
\bibitem [{\citenamefont {Steigman}\ and\ \citenamefont
  {Turner}(1985)}]{STEIGMAN1985375}%
  \BibitemOpen
  \bibfield  {author} {\bibinfo {author} {\bibfnamefont {G.}~\bibnamefont
  {Steigman}}\ and\ \bibinfo {author} {\bibfnamefont {M.~S.}\ \bibnamefont
  {Turner}},\ }\href {\doibase https://doi.org/10.1016/0550-3213(85)90537-1}
  {\bibfield  {journal} {\bibinfo  {journal} {Nuclear Physics B}\ }\textbf
  {\bibinfo {volume} {253}},\ \bibinfo {pages} {375} (\bibinfo {year}
  {1985})}\BibitemShut {NoStop}%
\bibitem [{\citenamefont {Roszkowski}\ \emph {et~al.}(2018)\citenamefont
  {Roszkowski}, \citenamefont {Sessolo},\ and\ \citenamefont
  {Trojanowski}}]{Roszkowski:2017nbc}%
  \BibitemOpen
  \bibfield  {author} {\bibinfo {author} {\bibfnamefont {L.}~\bibnamefont
  {Roszkowski}}, \bibinfo {author} {\bibfnamefont {E.~M.}\ \bibnamefont
  {Sessolo}}, \ and\ \bibinfo {author} {\bibfnamefont {S.}~\bibnamefont
  {Trojanowski}},\ }\href {\doibase 10.1088/1361-6633/aab913} {\bibfield
  {journal} {\bibinfo  {journal} {Rept. Prog. Phys.}\ }\textbf {\bibinfo
  {volume} {81}},\ \bibinfo {pages} {066201} (\bibinfo {year} {2018})},\
  \Eprint {http://arxiv.org/abs/1707.06277} {arXiv:1707.06277 [hep-ph]}
  \BibitemShut {NoStop}%
%%CITATION = ARXIV:1707.06277;%%
\bibitem [{\citenamefont {Schumann}(2019)}]{Schumann:2019eaa}%
  \BibitemOpen
  \bibfield  {author} {\bibinfo {author} {\bibfnamefont {M.}~\bibnamefont
  {Schumann}},\ }\href {\doibase 10.1088/1361-6471/ab2ea5} {\bibfield
  {journal} {\bibinfo  {journal} {J. Phys. G}\ }\textbf {\bibinfo {volume}
  {46}},\ \bibinfo {pages} {103003} (\bibinfo {year} {2019})},\ \Eprint
  {http://arxiv.org/abs/1903.03026} {arXiv:1903.03026 [astro-ph.CO]}
  \BibitemShut {NoStop}%
\bibitem [{\citenamefont {Battaglieri}\ \emph {et~al.}(2017)\citenamefont
  {Battaglieri} \emph {et~al.}}]{Battaglieri:2017aum}%
  \BibitemOpen
  \bibfield  {author} {\bibinfo {author} {\bibfnamefont {M.}~\bibnamefont
  {Battaglieri}} \emph {et~al.},\ }\href@noop {} {\enquote {\bibinfo {title}
  {{US Cosmic Visions: New Ideas in Dark Matter 2017: Community Report}},}\ }
  (\bibinfo {year} {2017}),\ \bibinfo {note} {\url{arXiv:1707.04591v1}},\
  \Eprint {http://arxiv.org/abs/1707.04591} {arXiv:1707.04591 [hep-ph]}
  \BibitemShut {NoStop}%
\bibitem [{\citenamefont {Krnjaic}\ \emph {et~al.}(2022)\citenamefont {Krnjaic}
  \emph {et~al.}}]{Krnjaic:2022ozp}%
  \BibitemOpen
  \bibfield  {author} {\bibinfo {author} {\bibfnamefont {G.}~\bibnamefont
  {Krnjaic}} \emph {et~al.},\ }\href@noop {} {\  (\bibinfo {year} {2022})},\
  \Eprint {http://arxiv.org/abs/2207.00597} {arXiv:2207.00597 [hep-ph]}
  \BibitemShut {NoStop}%
\bibitem [{\citenamefont {Fabbrichesi}\ \emph {et~al.}(2020)\citenamefont
  {Fabbrichesi}, \citenamefont {Gabrielli},\ and\ \citenamefont
  {Lanfranchi}}]{Fabbrichesi:2020wbt}%
  \BibitemOpen
  \bibfield  {author} {\bibinfo {author} {\bibfnamefont {M.}~\bibnamefont
  {Fabbrichesi}}, \bibinfo {author} {\bibfnamefont {E.}~\bibnamefont
  {Gabrielli}}, \ and\ \bibinfo {author} {\bibfnamefont {G.}~\bibnamefont
  {Lanfranchi}},\ }\href@noop {} {\emph {\bibinfo {title} {The Physics of the
  Dark Photon - A Primer}}}\ (\bibinfo  {publisher} {Springer International
  Publishing},\ \bibinfo {year} {2020})\BibitemShut {NoStop}%
\bibitem [{\citenamefont {Filippi}\ and\ \citenamefont
  {De~Napoli}(2020)}]{Filippi:2020kii}%
  \BibitemOpen
  \bibfield  {author} {\bibinfo {author} {\bibfnamefont {A.}~\bibnamefont
  {Filippi}}\ and\ \bibinfo {author} {\bibfnamefont {M.}~\bibnamefont
  {De~Napoli}},\ }\href {\doibase https://doi.org/10.1016/j.revip.2020.100042}
  {\bibfield  {journal} {\bibinfo  {journal} {Rev. Phys.}\ }\textbf {\bibinfo
  {volume} {5}},\ \bibinfo {pages} {100042} (\bibinfo {year} {2020})},\ \Eprint
  {http://arxiv.org/abs/2006.04640} {arXiv:2006.04640 [hep-ph]} \BibitemShut
  {NoStop}%
\bibitem [{\citenamefont {Liddle}(2003)}]{Liddle:1998ew}%
  \BibitemOpen
  \bibfield  {author} {\bibinfo {author} {\bibfnamefont {A.~R.}\ \bibnamefont
  {Liddle}},\ }\href@noop {} {\emph {\bibinfo {title} {{An introduction to
  modern cosmology}}}},\ edited by\ \bibinfo {editor} {\bibnamefont {Wiley}}\
  (\bibinfo {year} {2003})\BibitemShut {NoStop}%
%%CITATION = INSPIRE-484483;%%
\bibitem [{\citenamefont {Coy}\ \emph {et~al.}(2021)\citenamefont {Coy},
  \citenamefont {Hambye}, \citenamefont {Tytgat},\ and\ \citenamefont
  {Vanderheyden}}]{Coy:2021ann}%
  \BibitemOpen
  \bibfield  {author} {\bibinfo {author} {\bibfnamefont {R.}~\bibnamefont
  {Coy}}, \bibinfo {author} {\bibfnamefont {T.}~\bibnamefont {Hambye}},
  \bibinfo {author} {\bibfnamefont {M.~H.~G.}\ \bibnamefont {Tytgat}}, \ and\
  \bibinfo {author} {\bibfnamefont {L.}~\bibnamefont {Vanderheyden}},\ }\href
  {\doibase 10.1103/PhysRevD.104.055021} {\bibfield  {journal} {\bibinfo
  {journal} {Phys. Rev. D}\ }\textbf {\bibinfo {volume} {104}},\ \bibinfo
  {pages} {055021} (\bibinfo {year} {2021})},\ \Eprint
  {http://arxiv.org/abs/2105.01263} {arXiv:2105.01263 [hep-ph]} \BibitemShut
  {NoStop}%
\bibitem [{\citenamefont {Holdom}(1986)}]{Holdom:1985ag}%
  \BibitemOpen
  \bibfield  {author} {\bibinfo {author} {\bibfnamefont {B.}~\bibnamefont
  {Holdom}},\ }\href {\doibase 10.1016/0370-2693(86)91377-8} {\bibfield
  {journal} {\bibinfo  {journal} {Phys. Lett. B}\ }\textbf {\bibinfo {volume}
  {166}},\ \bibinfo {pages} {196} (\bibinfo {year} {1986})}\BibitemShut
  {NoStop}%
\bibitem [{\citenamefont {Batell}\ \emph {et~al.}(2009)\citenamefont {Batell},
  \citenamefont {Pospelov},\ and\ \citenamefont {Ritz}}]{Batell:2009di}%
  \BibitemOpen
  \bibfield  {author} {\bibinfo {author} {\bibfnamefont {B.}~\bibnamefont
  {Batell}}, \bibinfo {author} {\bibfnamefont {M.}~\bibnamefont {Pospelov}}, \
  and\ \bibinfo {author} {\bibfnamefont {A.}~\bibnamefont {Ritz}},\ }\href
  {\doibase 10.1103/PhysRevD.80.095024} {\bibfield  {journal} {\bibinfo
  {journal} {Phys. Rev. D}\ }\textbf {\bibinfo {volume} {80}},\ \bibinfo
  {pages} {095024} (\bibinfo {year} {2009})},\ \Eprint
  {http://arxiv.org/abs/0906.5614} {arXiv:0906.5614 [hep-ph]} \BibitemShut
  {NoStop}%
\bibitem [{\citenamefont {Tucker-Smith}\ and\ \citenamefont
  {Weiner}(2001)}]{Tucker-Smith:2001myb}%
  \BibitemOpen
  \bibfield  {author} {\bibinfo {author} {\bibfnamefont {D.}~\bibnamefont
  {Tucker-Smith}}\ and\ \bibinfo {author} {\bibfnamefont {N.}~\bibnamefont
  {Weiner}},\ }\href {\doibase 10.1103/PhysRevD.64.043502} {\bibfield
  {journal} {\bibinfo  {journal} {Phys. Rev. D}\ }\textbf {\bibinfo {volume}
  {64}},\ \bibinfo {pages} {043502} (\bibinfo {year} {2001})},\ \Eprint
  {http://arxiv.org/abs/hep-ph/0101138} {arXiv:hep-ph/0101138} \BibitemShut
  {NoStop}%
\bibitem [{\citenamefont {Madhavacheril}\ \emph {et~al.}(2014)\citenamefont
  {Madhavacheril}, \citenamefont {Sehgal},\ and\ \citenamefont
  {Slatyer}}]{Madhavacheril:2013cna}%
  \BibitemOpen
  \bibfield  {author} {\bibinfo {author} {\bibfnamefont {M.~S.}\ \bibnamefont
  {Madhavacheril}}, \bibinfo {author} {\bibfnamefont {N.}~\bibnamefont
  {Sehgal}}, \ and\ \bibinfo {author} {\bibfnamefont {T.~R.}\ \bibnamefont
  {Slatyer}},\ }\href {\doibase 10.1103/PhysRevD.89.103508} {\bibfield
  {journal} {\bibinfo  {journal} {Phys. Rev. D}\ }\textbf {\bibinfo {volume}
  {89}},\ \bibinfo {pages} {103508} (\bibinfo {year} {2014})},\ \Eprint
  {http://arxiv.org/abs/1310.3815} {arXiv:1310.3815 [astro-ph.CO]} \BibitemShut
  {NoStop}%
\bibitem [{\citenamefont {\r{A}kesson}\ \emph {et~al.}(2018)\citenamefont
  {\r{A}kesson} \emph {et~al.}}]{LDMX:2018cma}%
  \BibitemOpen
  \bibfield  {author} {\bibinfo {author} {\bibfnamefont {T.}~\bibnamefont
  {\r{A}kesson}} \emph {et~al.} (\bibinfo {collaboration} {LDMX}),\ }\href@noop
  {} {\  (\bibinfo {year} {2018})},\ \Eprint {http://arxiv.org/abs/1808.05219}
  {arXiv:1808.05219 [hep-ex]} \BibitemShut {NoStop}%
\bibitem [{\citenamefont {Beacham}\ \emph {et~al.}(2020)\citenamefont {Beacham}
  \emph {et~al.}}]{Beacham:2019nyx}%
  \BibitemOpen
  \bibfield  {author} {\bibinfo {author} {\bibfnamefont {J.}~\bibnamefont
  {Beacham}} \emph {et~al.},\ }\href {\doibase 10.1088/1361-6471/ab4cd2}
  {\bibfield  {journal} {\bibinfo  {journal} {J. Phys. G}\ }\textbf {\bibinfo
  {volume} {47}},\ \bibinfo {pages} {010501} (\bibinfo {year} {2020})},\
  \Eprint {http://arxiv.org/abs/1901.09966} {arXiv:1901.09966 [hep-ex]}
  \BibitemShut {NoStop}%
\bibitem [{\citenamefont {Batell}\ \emph {et~al.}(2014)\citenamefont {Batell},
  \citenamefont {Essig},\ and\ \citenamefont {Surujon}}]{Batell:2014mga}%
  \BibitemOpen
  \bibfield  {author} {\bibinfo {author} {\bibfnamefont {B.}~\bibnamefont
  {Batell}}, \bibinfo {author} {\bibfnamefont {R.}~\bibnamefont {Essig}}, \
  and\ \bibinfo {author} {\bibfnamefont {Z.}~\bibnamefont {Surujon}},\ }\href
  {\doibase 10.1103/PhysRevLett.113.171802} {\bibfield  {journal} {\bibinfo
  {journal} {Phys. Rev. Lett.}\ }\textbf {\bibinfo {volume} {113}},\ \bibinfo
  {pages} {171802} (\bibinfo {year} {2014})},\ \Eprint
  {http://arxiv.org/abs/1406.2698} {arXiv:1406.2698 [hep-ph]} \BibitemShut
  {NoStop}%
\bibitem [{\citenamefont {Andreev}\ \emph {et~al.}(2021)\citenamefont {Andreev}
  \emph {et~al.}}]{Andreev:2021fzd}%
  \BibitemOpen
  \bibfield  {author} {\bibinfo {author} {\bibfnamefont {Y.~M.}\ \bibnamefont
  {Andreev}} \emph {et~al.},\ }\href {\doibase 10.1103/PhysRevD.104.L091701}
  {\bibfield  {journal} {\bibinfo  {journal} {Phys. Rev. D}\ }\textbf {\bibinfo
  {volume} {104}},\ \bibinfo {pages} {L091701} (\bibinfo {year} {2021})},\
  \Eprint {http://arxiv.org/abs/2108.04195} {arXiv:2108.04195 [hep-ex]}
  \BibitemShut {NoStop}%
\bibitem [{\citenamefont {Bond\'\i{}}(2017)}]{Bondi:2017gul}%
  \BibitemOpen
  \bibfield  {author} {\bibinfo {author} {\bibfnamefont {M.}~\bibnamefont
  {Bond\'\i{}}} (\bibinfo {collaboration} {BDX}),\ }\href {\doibase
  10.1051/epjconf/201714201005} {\bibfield  {journal} {\bibinfo  {journal} {EPJ
  Web Conf.}\ }\textbf {\bibinfo {volume} {142}},\ \bibinfo {pages} {01005}
  (\bibinfo {year} {2017})}\BibitemShut {NoStop}%
\bibitem [{\citenamefont {Izaguirre}\ \emph {et~al.}(2013)\citenamefont
  {Izaguirre}, \citenamefont {Krnjaic}, \citenamefont {Schuster},\ and\
  \citenamefont {Toro}}]{Izaguirre:2013uxa}%
  \BibitemOpen
  \bibfield  {author} {\bibinfo {author} {\bibfnamefont {E.}~\bibnamefont
  {Izaguirre}}, \bibinfo {author} {\bibfnamefont {G.}~\bibnamefont {Krnjaic}},
  \bibinfo {author} {\bibfnamefont {P.}~\bibnamefont {Schuster}}, \ and\
  \bibinfo {author} {\bibfnamefont {N.}~\bibnamefont {Toro}},\ }\href {\doibase
  10.1103/PhysRevD.88.114015} {\bibfield  {journal} {\bibinfo  {journal} {Phys.
  Rev. D}\ }\textbf {\bibinfo {volume} {88}},\ \bibinfo {pages} {114015}
  (\bibinfo {year} {2013})},\ \Eprint {http://arxiv.org/abs/1307.6554}
  {arXiv:1307.6554 [hep-ph]} \BibitemShut {NoStop}%
\bibitem [{\citenamefont {Marsicano}\ \emph
  {et~al.}(2018{\natexlab{a}})\citenamefont {Marsicano}, \citenamefont
  {Battaglieri}, \citenamefont {Bond\'\i{}}, \citenamefont {Carvajal},
  \citenamefont {Celentano}, \citenamefont {De~Napoli}, \citenamefont
  {De~Vita}, \citenamefont {Nardi}, \citenamefont {Raggi},\ and\ \citenamefont
  {Valente}}]{Marsicano:2018glj}%
  \BibitemOpen
  \bibfield  {author} {\bibinfo {author} {\bibfnamefont {L.}~\bibnamefont
  {Marsicano}}, \bibinfo {author} {\bibfnamefont {M.}~\bibnamefont
  {Battaglieri}}, \bibinfo {author} {\bibfnamefont {M.}~\bibnamefont
  {Bond\'\i{}}}, \bibinfo {author} {\bibfnamefont {C.~R.}\ \bibnamefont
  {Carvajal}}, \bibinfo {author} {\bibfnamefont {A.}~\bibnamefont {Celentano}},
  \bibinfo {author} {\bibfnamefont {M.}~\bibnamefont {De~Napoli}}, \bibinfo
  {author} {\bibfnamefont {R.}~\bibnamefont {De~Vita}}, \bibinfo {author}
  {\bibfnamefont {E.}~\bibnamefont {Nardi}}, \bibinfo {author} {\bibfnamefont
  {M.}~\bibnamefont {Raggi}}, \ and\ \bibinfo {author} {\bibfnamefont
  {P.}~\bibnamefont {Valente}},\ }\href {\doibase
  10.1103/PhysRevLett.121.041802} {\bibfield  {journal} {\bibinfo  {journal}
  {Phys. Rev. Lett.}\ }\textbf {\bibinfo {volume} {121}},\ \bibinfo {pages}
  {041802} (\bibinfo {year} {2018}{\natexlab{a}})},\ \Eprint
  {http://arxiv.org/abs/1807.05884} {arXiv:1807.05884 [hep-ex]} \BibitemShut
  {NoStop}%
\bibitem [{\citenamefont {Marsicano}\ \emph
  {et~al.}(2018{\natexlab{b}})\citenamefont {Marsicano}, \citenamefont
  {Battaglieri}, \citenamefont {Bond\'{\i}}, \citenamefont {Carvajal},
  \citenamefont {Celentano}, \citenamefont {De~Napoli}, \citenamefont
  {De~Vita}, \citenamefont {Nardi}, \citenamefont {Raggi},\ and\ \citenamefont
  {Valente}}]{Marsicano:2018krp}%
  \BibitemOpen
  \bibfield  {author} {\bibinfo {author} {\bibfnamefont {L.}~\bibnamefont
  {Marsicano}}, \bibinfo {author} {\bibfnamefont {M.}~\bibnamefont
  {Battaglieri}}, \bibinfo {author} {\bibfnamefont {M.}~\bibnamefont
  {Bond\'{\i}}}, \bibinfo {author} {\bibfnamefont {C.~D.~R.}\ \bibnamefont
  {Carvajal}}, \bibinfo {author} {\bibfnamefont {A.}~\bibnamefont {Celentano}},
  \bibinfo {author} {\bibfnamefont {M.}~\bibnamefont {De~Napoli}}, \bibinfo
  {author} {\bibfnamefont {R.}~\bibnamefont {De~Vita}}, \bibinfo {author}
  {\bibfnamefont {E.}~\bibnamefont {Nardi}}, \bibinfo {author} {\bibfnamefont
  {M.}~\bibnamefont {Raggi}}, \ and\ \bibinfo {author} {\bibfnamefont
  {P.}~\bibnamefont {Valente}},\ }\href {\doibase 10.1103/PhysRevD.98.015031}
  {\bibfield  {journal} {\bibinfo  {journal} {Phys. Rev. D}\ }\textbf {\bibinfo
  {volume} {98}},\ \bibinfo {pages} {015031} (\bibinfo {year}
  {2018}{\natexlab{b}})},\ \Eprint {http://arxiv.org/abs/1802.03794}
  {arXiv:1802.03794 [hep-ex]} \BibitemShut {NoStop}%
\bibitem [{\citenamefont {Battaglieri}\ \emph {et~al.}(2019)\citenamefont
  {Battaglieri} \emph {et~al.}}]{BDX:2019afh}%
  \BibitemOpen
  \bibfield  {author} {\bibinfo {author} {\bibfnamefont {M.}~\bibnamefont
  {Battaglieri}} \emph {et~al.} (\bibinfo {collaboration} {BDX}),\ }\href@noop
  {} {\  (\bibinfo {year} {2019})},\ \Eprint {http://arxiv.org/abs/1910.03532}
  {arXiv:1910.03532 [physics.ins-det]} \BibitemShut {NoStop}%
\bibitem [{\citenamefont {Aubert}\ \emph {et~al.}(2002)\citenamefont {Aubert}
  \emph {et~al.}}]{BaBar:2001yhh}%
  \BibitemOpen
  \bibfield  {author} {\bibinfo {author} {\bibfnamefont {B.}~\bibnamefont
  {Aubert}} \emph {et~al.} (\bibinfo {collaboration} {BaBar}),\ }\href
  {\doibase 10.1016/S0168-9002(01)02012-5} {\bibfield  {journal} {\bibinfo
  {journal} {Nucl. Instrum. Meth. A}\ }\textbf {\bibinfo {volume} {479}},\
  \bibinfo {pages} {1} (\bibinfo {year} {2002})},\ \Eprint
  {http://arxiv.org/abs/hep-ex/0105044} {arXiv:hep-ex/0105044} \BibitemShut
  {NoStop}%
\bibitem [{\citenamefont {Battaglieri}\ \emph {et~al.}(2021)\citenamefont
  {Battaglieri} \emph {et~al.}}]{Battaglieri:2020lds}%
  \BibitemOpen
  \bibfield  {author} {\bibinfo {author} {\bibfnamefont {M.}~\bibnamefont
  {Battaglieri}} \emph {et~al.},\ }\href {\doibase
  10.1140/epjc/s10052-021-08957-5} {\bibfield  {journal} {\bibinfo  {journal}
  {Eur. Phys. J. C}\ }\textbf {\bibinfo {volume} {81}},\ \bibinfo {pages} {164}
  (\bibinfo {year} {2021})},\ \Eprint {http://arxiv.org/abs/2011.10532}
  {arXiv:2011.10532 [physics.ins-det]} \BibitemShut {NoStop}%
\bibitem [{COD(2018)}]{CODA}%
  \BibitemOpen
  \href {https://coda.jlab.org/drupal/} {\enquote {\bibinfo {title} {{JLab
  CODA}},}\ } (\bibinfo {year} {2018})\BibitemShut {NoStop}%
\bibitem [{\citenamefont {Dalesio}\ \emph {et~al.}(1994)\citenamefont
  {Dalesio}, \citenamefont {Hill}, \citenamefont {Kraimer}, \citenamefont
  {Lewis}, \citenamefont {Murray}, \citenamefont {Hunt}, \citenamefont
  {Watson}, \citenamefont {Clausen},\ and\ \citenamefont
  {Dalesio}}]{Dalesio:1994qp}%
  \BibitemOpen
  \bibfield  {author} {\bibinfo {author} {\bibfnamefont {L.~R.}\ \bibnamefont
  {Dalesio}}, \bibinfo {author} {\bibfnamefont {J.~O.}\ \bibnamefont {Hill}},
  \bibinfo {author} {\bibfnamefont {M.}~\bibnamefont {Kraimer}}, \bibinfo
  {author} {\bibfnamefont {S.}~\bibnamefont {Lewis}}, \bibinfo {author}
  {\bibfnamefont {D.}~\bibnamefont {Murray}}, \bibinfo {author} {\bibfnamefont
  {S.}~\bibnamefont {Hunt}}, \bibinfo {author} {\bibfnamefont {W.}~\bibnamefont
  {Watson}}, \bibinfo {author} {\bibfnamefont {M.}~\bibnamefont {Clausen}}, \
  and\ \bibinfo {author} {\bibfnamefont {J.}~\bibnamefont {Dalesio}},\ }\href
  {\doibase 10.1016/0168-9002(94)91493-1} {\bibfield  {journal} {\bibinfo
  {journal} {Nucl. Instrum. Meth. A}\ }\textbf {\bibinfo {volume} {352}},\
  \bibinfo {pages} {179} (\bibinfo {year} {1994})}\BibitemShut {NoStop}%
\bibitem [{\citenamefont {Böhlen}\ \emph {et~al.}(2014)\citenamefont
  {Böhlen}, \citenamefont {Cerutti}, \citenamefont {Chin}, \citenamefont
  {Fassò}, \citenamefont {Ferrari}, \citenamefont {Ortega}, \citenamefont
  {Mairani}, \citenamefont {Sala}, \citenamefont {Smirnov},\ and\ \citenamefont
  {Vlachoudis}}]{Bohlen:2014buj}%
  \BibitemOpen
  \bibfield  {author} {\bibinfo {author} {\bibfnamefont {T.~T.}\ \bibnamefont
  {Böhlen}}, \bibinfo {author} {\bibfnamefont {F.}~\bibnamefont {Cerutti}},
  \bibinfo {author} {\bibfnamefont {M.~P.~W.}\ \bibnamefont {Chin}}, \bibinfo
  {author} {\bibfnamefont {A.}~\bibnamefont {Fassò}}, \bibinfo {author}
  {\bibfnamefont {A.}~\bibnamefont {Ferrari}}, \bibinfo {author} {\bibfnamefont
  {P.~G.}\ \bibnamefont {Ortega}}, \bibinfo {author} {\bibfnamefont
  {A.}~\bibnamefont {Mairani}}, \bibinfo {author} {\bibfnamefont {P.~R.}\
  \bibnamefont {Sala}}, \bibinfo {author} {\bibfnamefont {G.}~\bibnamefont
  {Smirnov}}, \ and\ \bibinfo {author} {\bibfnamefont {V.}~\bibnamefont
  {Vlachoudis}},\ }\href {\doibase 10.1016/j.nds.2014.07.049} {\bibfield
  {journal} {\bibinfo  {journal} {Nucl. Data Sheets}\ }\textbf {\bibinfo
  {volume} {120}},\ \bibinfo {pages} {211} (\bibinfo {year}
  {2014})}\BibitemShut {NoStop}%
%%CITATION = NDTSB,120,211;%%
\bibitem [{\citenamefont {Ferrari}\ \emph {et~al.}()\citenamefont {Ferrari},
  \citenamefont {Sala}, \citenamefont {Fasso},\ and\ \citenamefont
  {Ranft}}]{Ferrari:2005zk}%
  \BibitemOpen
  \bibfield  {author} {\bibinfo {author} {\bibfnamefont {A.}~\bibnamefont
  {Ferrari}}, \bibinfo {author} {\bibfnamefont {P.~R.}\ \bibnamefont {Sala}},
  \bibinfo {author} {\bibfnamefont {A.}~\bibnamefont {Fasso}}, \ and\ \bibinfo
  {author} {\bibfnamefont {J.}~\bibnamefont {Ranft}},\ }\href {\doibase
  10.2172/877507} {\emph {\bibinfo {title} {FLUKA: A multi-particle transport
  code (Program version 2005)}}}\BibitemShut {NoStop}%
\bibitem [{\citenamefont {Kharashvili}(2016)}]{Kharas}%
  \BibitemOpen
  \bibfield  {author} {\bibinfo {author} {\bibfnamefont {M.}~\bibnamefont
  {Kharashvili}},\ }\href@noop {} {\emph {\bibinfo {title}
  {{JLAB-TN-16-048}}}},\ \bibinfo {type} {Tech. Rep.}\ (\bibinfo {year}
  {2016})\BibitemShut {NoStop}%
\bibitem [{\citenamefont {Alwall}\ \emph {et~al.}(2007)\citenamefont {Alwall},
  \citenamefont {Demin}, \citenamefont {de~Visscher}, \citenamefont {Frederix},
  \citenamefont {Herquet}, \citenamefont {Maltoni}, \citenamefont {Plehn},
  \citenamefont {Rainwater},\ and\ \citenamefont {Stelzer}}]{Alwall:2007st}%
  \BibitemOpen
  \bibfield  {author} {\bibinfo {author} {\bibfnamefont {J.}~\bibnamefont
  {Alwall}}, \bibinfo {author} {\bibfnamefont {P.}~\bibnamefont {Demin}},
  \bibinfo {author} {\bibfnamefont {S.}~\bibnamefont {de~Visscher}}, \bibinfo
  {author} {\bibfnamefont {R.}~\bibnamefont {Frederix}}, \bibinfo {author}
  {\bibfnamefont {M.}~\bibnamefont {Herquet}}, \bibinfo {author} {\bibfnamefont
  {F.}~\bibnamefont {Maltoni}}, \bibinfo {author} {\bibfnamefont
  {T.}~\bibnamefont {Plehn}}, \bibinfo {author} {\bibfnamefont {D.~L.}\
  \bibnamefont {Rainwater}}, \ and\ \bibinfo {author} {\bibfnamefont
  {T.}~\bibnamefont {Stelzer}},\ }\href {\doibase
  10.1088/1126-6708/2007/09/028} {\bibfield  {journal} {\bibinfo  {journal}
  {JHEP}\ }\textbf {\bibinfo {volume} {09}},\ \bibinfo {pages} {028} (\bibinfo
  {year} {2007})},\ \Eprint {http://arxiv.org/abs/0706.2334} {arXiv:0706.2334
  [hep-ph]} \BibitemShut {NoStop}%
\bibitem [{\citenamefont {Bjorken}\ \emph {et~al.}(2009)\citenamefont
  {Bjorken}, \citenamefont {Essig}, \citenamefont {Schuster},\ and\
  \citenamefont {Toro}}]{Bjorken:2009mm}%
  \BibitemOpen
  \bibfield  {author} {\bibinfo {author} {\bibfnamefont {J.~D.}\ \bibnamefont
  {Bjorken}}, \bibinfo {author} {\bibfnamefont {R.}~\bibnamefont {Essig}},
  \bibinfo {author} {\bibfnamefont {P.}~\bibnamefont {Schuster}}, \ and\
  \bibinfo {author} {\bibfnamefont {N.}~\bibnamefont {Toro}},\ }\href {\doibase
  10.1103/PhysRevD.80.075018} {\bibfield  {journal} {\bibinfo  {journal} {Phys.
  Rev. D}\ }\textbf {\bibinfo {volume} {80}},\ \bibinfo {pages} {075018}
  (\bibinfo {year} {2009})},\ \Eprint {http://arxiv.org/abs/0906.0580}
  {arXiv:0906.0580 [hep-ph]} \BibitemShut {NoStop}%
\bibitem [{\citenamefont {Andreopoulos}\ \emph {et~al.}(2010)\citenamefont
  {Andreopoulos} \emph {et~al.}}]{Andreopoulos:2009rq}%
  \BibitemOpen
  \bibfield  {author} {\bibinfo {author} {\bibfnamefont {C.}~\bibnamefont
  {Andreopoulos}} \emph {et~al.},\ }\href {\doibase 10.1016/j.nima.2009.12.009}
  {\bibfield  {journal} {\bibinfo  {journal} {Nucl. Instrum. Meth. A}\ }\textbf
  {\bibinfo {volume} {614}},\ \bibinfo {pages} {87} (\bibinfo {year} {2010})},\
  \Eprint {http://arxiv.org/abs/0905.2517} {arXiv:0905.2517 [hep-ph]}
  \BibitemShut {NoStop}%
\bibitem [{\citenamefont {Andreopoulos}\ \emph {et~al.}(2015)\citenamefont
  {Andreopoulos}, \citenamefont {Barry}, \citenamefont {Dytman}, \citenamefont
  {Gallagher}, \citenamefont {Golan}, \citenamefont {Hatcher}, \citenamefont
  {Perdue},\ and\ \citenamefont {Yarba}}]{Andreopoulos:2015wxa}%
  \BibitemOpen
  \bibfield  {author} {\bibinfo {author} {\bibfnamefont {C.}~\bibnamefont
  {Andreopoulos}}, \bibinfo {author} {\bibfnamefont {C.}~\bibnamefont {Barry}},
  \bibinfo {author} {\bibfnamefont {S.}~\bibnamefont {Dytman}}, \bibinfo
  {author} {\bibfnamefont {H.}~\bibnamefont {Gallagher}}, \bibinfo {author}
  {\bibfnamefont {T.}~\bibnamefont {Golan}}, \bibinfo {author} {\bibfnamefont
  {R.}~\bibnamefont {Hatcher}}, \bibinfo {author} {\bibfnamefont
  {G.}~\bibnamefont {Perdue}}, \ and\ \bibinfo {author} {\bibfnamefont
  {J.}~\bibnamefont {Yarba}},\ }\href@noop {} {\enquote {\bibinfo {title} {{The
  GENIE Neutrino Monte Carlo Generator: Physics and User Manual}},}\ }
  (\bibinfo {year} {2015}),\ \Eprint {http://arxiv.org/abs/1510.05494}
  {arXiv:1510.05494 [hep-ph]} \BibitemShut {NoStop}%
\bibitem [{\citenamefont {Berger}(2018)}]{Berger:2018urf}%
  \BibitemOpen
  \bibfield  {author} {\bibinfo {author} {\bibfnamefont {J.}~\bibnamefont
  {Berger}},\ }\href@noop {} {\  (\bibinfo {year} {2018})},\ \Eprint
  {http://arxiv.org/abs/1812.05616} {arXiv:1812.05616 [hep-ph]} \BibitemShut
  {NoStop}%
\bibitem [{\citenamefont {Gillessen}\ and\ \citenamefont
  {Harney}(2005)}]{Gillessen:2004pm}%
  \BibitemOpen
  \bibfield  {author} {\bibinfo {author} {\bibfnamefont {S.}~\bibnamefont
  {Gillessen}}\ and\ \bibinfo {author} {\bibfnamefont {H.~L.}\ \bibnamefont
  {Harney}},\ }\href {\doibase 10.1051/0004-6361:20035839} {\bibfield
  {journal} {\bibinfo  {journal} {Astron. Astrophys.}\ }\textbf {\bibinfo
  {volume} {430}},\ \bibinfo {pages} {355} (\bibinfo {year} {2005})},\ \Eprint
  {http://arxiv.org/abs/astro-ph/0411660} {arXiv:astro-ph/0411660} \BibitemShut
  {NoStop}%
\bibitem [{\citenamefont {de~Romeri}\ \emph {et~al.}(2020)\citenamefont
  {de~Romeri}, \citenamefont {Kelly},\ and\ \citenamefont
  {Machado}}]{deRomeri:2020kno}%
  \BibitemOpen
  \bibfield  {author} {\bibinfo {author} {\bibfnamefont {V.}~\bibnamefont
  {de~Romeri}}, \bibinfo {author} {\bibfnamefont {K.~J.}\ \bibnamefont
  {Kelly}}, \ and\ \bibinfo {author} {\bibfnamefont {P.~A.~N.}\ \bibnamefont
  {Machado}},\ }\href {\doibase 10.1088/1742-6596/1468/1/012061} {\bibfield
  {journal} {\bibinfo  {journal} {J. Phys. Conf. Ser.}\ }\textbf {\bibinfo
  {volume} {1468}},\ \bibinfo {pages} {012061} (\bibinfo {year}
  {2020})}\BibitemShut {NoStop}%
\bibitem [{\citenamefont {Moneta}\ \emph {et~al.}(2010)\citenamefont {Moneta},
  \citenamefont {Belasco}, \citenamefont {Cranmer}, \citenamefont {Kreiss},
  \citenamefont {Lazzaro}, \citenamefont {Piparo}, \citenamefont {Schott},
  \citenamefont {Verkerke},\ and\ \citenamefont {Wolf}}]{Moneta:2010pm}%
  \BibitemOpen
  \bibfield  {author} {\bibinfo {author} {\bibfnamefont {L.}~\bibnamefont
  {Moneta}}, \bibinfo {author} {\bibfnamefont {K.}~\bibnamefont {Belasco}},
  \bibinfo {author} {\bibfnamefont {K.~S.}\ \bibnamefont {Cranmer}}, \bibinfo
  {author} {\bibfnamefont {S.}~\bibnamefont {Kreiss}}, \bibinfo {author}
  {\bibfnamefont {A.}~\bibnamefont {Lazzaro}}, \bibinfo {author} {\bibfnamefont
  {D.}~\bibnamefont {Piparo}}, \bibinfo {author} {\bibfnamefont
  {G.}~\bibnamefont {Schott}}, \bibinfo {author} {\bibfnamefont
  {W.}~\bibnamefont {Verkerke}}, \ and\ \bibinfo {author} {\bibfnamefont
  {M.}~\bibnamefont {Wolf}},\ }\href {\doibase 10.22323/1.093.0057} {\bibfield
  {journal} {\bibinfo  {journal} {PoS}\ }\textbf {\bibinfo {volume}
  {ACAT2010}},\ \bibinfo {pages} {057} (\bibinfo {year} {2010})},\ \Eprint
  {http://arxiv.org/abs/1009.1003} {arXiv:1009.1003 [physics.data-an]}
  \BibitemShut {NoStop}%
\bibitem [{\citenamefont {Lees}\ \emph {et~al.}(2017)\citenamefont {Lees} \emph
  {et~al.}}]{BaBar:2017tiz}%
  \BibitemOpen
  \bibfield  {author} {\bibinfo {author} {\bibfnamefont {J.~P.}\ \bibnamefont
  {Lees}} \emph {et~al.} (\bibinfo {collaboration} {BaBar}),\ }\href {\doibase
  10.1103/PhysRevLett.119.131804} {\bibfield  {journal} {\bibinfo  {journal}
  {Phys. Rev. Lett.}\ }\textbf {\bibinfo {volume} {119}},\ \bibinfo {pages}
  {131804} (\bibinfo {year} {2017})},\ \Eprint
  {http://arxiv.org/abs/1702.03327} {arXiv:1702.03327 [hep-ex]} \BibitemShut
  {NoStop}%
\bibitem [{\citenamefont {Banerjee}\ \emph {et~al.}(2019)\citenamefont
  {Banerjee} \emph {et~al.}}]{Banerjee:2019pds}%
  \BibitemOpen
  \bibfield  {author} {\bibinfo {author} {\bibfnamefont {D.}~\bibnamefont
  {Banerjee}} \emph {et~al.},\ }\href {\doibase 10.1103/PhysRevLett.123.121801}
  {\bibfield  {journal} {\bibinfo  {journal} {Phys. Rev. Lett.}\ }\textbf
  {\bibinfo {volume} {123}},\ \bibinfo {pages} {121801} (\bibinfo {year}
  {2019})},\ \Eprint {http://arxiv.org/abs/1906.00176} {arXiv:1906.00176
  [hep-ex]} \BibitemShut {NoStop}%
\bibitem [{\citenamefont {Aguilar-Arevalo}\ \emph {et~al.}(2018)\citenamefont
  {Aguilar-Arevalo} \emph {et~al.}}]{MiniBooNEDM:2018cxm}%
  \BibitemOpen
  \bibfield  {author} {\bibinfo {author} {\bibfnamefont {A.~A.}\ \bibnamefont
  {Aguilar-Arevalo}} \emph {et~al.} (\bibinfo {collaboration} {MiniBooNE DM}),\
  }\href {\doibase 10.1103/PhysRevD.98.112004} {\bibfield  {journal} {\bibinfo
  {journal} {Phys. Rev. D}\ }\textbf {\bibinfo {volume} {98}},\ \bibinfo
  {pages} {112004} (\bibinfo {year} {2018})},\ \Eprint
  {http://arxiv.org/abs/1807.06137} {arXiv:1807.06137 [hep-ex]} \BibitemShut
  {NoStop}%
\bibitem [{\citenamefont {Akimov}\ \emph {et~al.}(2021)\citenamefont {Akimov}
  \emph {et~al.}}]{COHERENT:2021pvd}%
  \BibitemOpen
  \bibfield  {author} {\bibinfo {author} {\bibfnamefont {D.}~\bibnamefont
  {Akimov}} \emph {et~al.} (\bibinfo {collaboration} {COHERENT}),\ }\href@noop
  {} {\  (\bibinfo {year} {2021})},\ \Eprint {http://arxiv.org/abs/2110.11453}
  {arXiv:2110.11453 [hep-ex]} \BibitemShut {NoStop}%
\end{thebibliography}%

\end{document}